\documentclass[twocolumn,tighten,times]{aastex631}
\usepackage{xcolor}
\usepackage{apjfonts}

\newcommand{\cajwl}{$Ca_{\rm JWL}$}
\newcommand{\chjwl}{$ch_{\rm JWL}$}
\newcommand{\nhjwl}{$nh_{\rm JWL}$}
\newcommand{\cnjwl}{$cn_{\rm JWL}$}
\newcommand{\dvbump}{$\Delta V_{\rm bump}$}

\newcommand{\nthree}{$n$(FG):$n$(IG):$n$(SG)}
\newcommand{\nrgb}{$n$(FG):$n$(SG)}

\newcommand{\str}{Str\"omgren}
\newcommand{\vbump}{$V_{\rm bump}$}

\newcommand{\vvhbmag}{$-$2.5 mag $\leq$ $V - V_{\rm HB}$ $\leq$ 2.0 mag}
\newcommand{\cnwave}{$\lambda$3883}

\newcommand{\dy}{$\Delta Y$}
\newcommand{\hkjwl}{$hk_{\rm JWL}$}

\newcommand{\teff}{$T_{\rm eff}$}

\newcommand{\fehhk}{[Fe/H]$_{hk}$}
\newcommand{\cfech}{[C/Fe]$_{ch}$}
\newcommand{\nfenh}{[N/Fe]$_{nh}$}
\newcommand{\pcfe}{$\parallel$\,[C/Fe]$_{ch}$}
\newcommand{\pnfe}{$\parallel$\,[N/Fe]$_{nh}$}
\newcommand{\dcfenfe}{$\parallel$\,[C/Fe]$_{ch}$ $-$ $\parallel$\,[N/Fe]$_{nh}$}
\newcommand{\cfe}{[C/Fe]}
\newcommand{\nfe}{[N/Fe]}
\newcommand{\feh}{[Fe/H]}

\newcommand{\dfeh}{$\Delta$[Fe/H]}
\newcommand{\dnfe}{$\Delta$[N/Fe]}
\newcommand{\dcfe}{$\Delta$[C/Fe]}
\newcommand{\msun}{$M_{\rm \odot}$}
\newcommand{\ebv}{$E(B-V)$}

\newcommand{\cubi}{$C_{\rm UBI}$}
\newcommand{\pcubi}{$\parallel\,C_{\rm UBI}$}

\newcommand{\linemake}{\texttt{Linemake}}
\newcommand{\moogscat}{\texttt{MOOG\_SCAT}}
\newcommand{\atlas}{\texttt{ATLAS12}}

\shortauthors{Lee}

\begin{document}

\title{A Comparative Study Between M30 and M92 : M92 Is A Merger Remnant With A Large Helium Enhancement}

\author[0000-0002-2122-3030]{Jae-Woo Lee}
\affiliation{Department of Physics and Astronomy, Sejong University, 209 Neungdong-ro, Gwangjin-Gu, Seoul 05006, Republic of Korea,
jaewoolee@sejong.ac.kr, jaewoolee@sejong.edu}

\begin{abstract}
We perform a comparative study of the ex--situ second--parameter pair globular clusters (GCs) M30 and M92, having similar metallicities but different horizontal branch morphologies. We obtain the similar mean primordial carbon abundances for both clusters. However, M92 shows a large dispersion in carbon due to a more extended C--N anticorrelation, while M30 exhibits a higher primordial nitrogen abundance, suggesting that they have different chemical enrichment histories.
Our new results confirm our previous result that M92 is a metal--complex GC showing a bimodal metallicity distribution. We also find that the metal--rich group of stars in M92 shows a helium enhancement as large as $\Delta Y$ $\sim$ 0.05 from the red giant branch bump (RGBB) $V$ magnitudes, which can also be supported by (i) a lack of bright RGB stars, (ii) synthetic evolutionary HB population models and (iii) the more extended spatial distribution due to different degree of the diffusion process from their lower masses. We reinterpret the [Eu/Fe] measurements by other, finding that the two metallicity groups of stars in M92 have significantly different [Eu/Fe] abundances with small scatters. This strongly suggests that they formed independently out of well mixed interstellar media in different environments. We suggest that M92 is a more complex system than a normal GC, most likely a merger remnant of two GCs or a even more complex system. In Appendix, we address the problems with the recently developed color--temperature relations and the usage of broadband photometry in the populational taggings.
\end{abstract}


\section{INTRODUCTION}
Galactic globular clusters (GCs) had played critical roles in stellar and Galactic astronomy for several decades. Due to the seminal discovery of multiple stellar populations (MSPs) in GCs \citep{dercole08, carretta09, lee09, milone17, gratton19, cassisi20}, the present situation may not be as lucid as before, since the homogeneous chemical composition of a given GC is no longer valid. Nonetheless, they are still the unique object to calibrate the theoretical models for the structure and evolution of low--mass stars with various chemical compositions \citep[e.g., see][]{cassisi13, vandenberg22}.

Galactic GCs have also been extensively studied to  figure out the formation and evolution of our Galaxy. During the past several decades, not only the in--situ but also the ex--situ GCs have been identified \citep[e.g.,][]{sz, lee99}, which can nicely be interpreted with a cold dark matter cosmology that predicts a hierarchical structure formation in the universe.

In this study, we perform a comparative study of two ex--situ metal--poor GCs, M92 (NGC\,6341) and M30 (NGC\,7099), which have long been considered as standard metal--poor GCs. Recent studies of M92 discovered a stellar stream related to M92 \citep{sollima20, thomas20}. Furthermore, \citet{thomas20} argued that M92 could be a GC that was brought into our Galaxy by a dwarf galaxy or a remnant nucleus of the progenitor dwarf galaxy.\footnote{Note that M92 is known to be associated to Gaia--Enceladus \citep{forbes20, callingham22}.} In our previous study, we discussed that M92 is a metal--complex GC with an atypical primordial population \citep{lee23a}, in sharp contrast to a long-held belief of a prototypical mono--metallic metal--poor GC. As \citet{bekki16} proposed, for example, GCs in dwarf galaxies with slightly different formation epochs can merge into a metal--complex GC, such as M92. Interestingly, \citet{kirby23} found the $r$--process elemental abundance variations only in the first generation (FG)\footnote{In our work, the FG denotes GC stars having the similar chemical compositions as the Galactic field stars with the same metallicity, while the second generation (SG) of stars showing the chemical compositions that experienced proton capture processes at high temperature.} of stars, apparently constraining temporal separation between the FG and the SG of stars. On the other hand, M30 is known to be a member of the Gaia--Enceladus system, that merged into our Galaxy about 10 Gyr ago leading to the dynamical heating of the precursor of the Galactic thick disk \citep{helmi18}.

\begin{figure}
\epsscale{1.1}
\figurenum{1}
\plotone{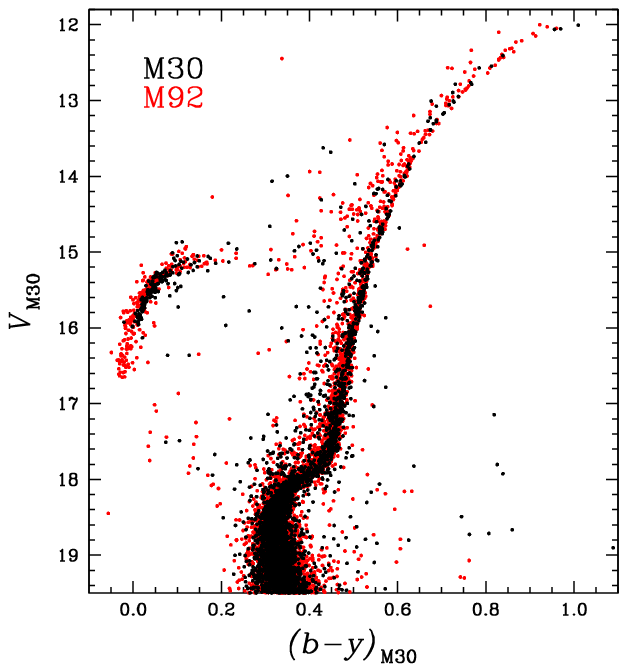}
\caption{A comparison of M30 and M92 CMDs. The $V$ magnitude and $(b-y)$ color for M92 are adjusted to match with those of M30.}\label{fig:by}
\end{figure}

\begin{figure*}
\epsscale{1.1}
\figurenum{2}
\plotone{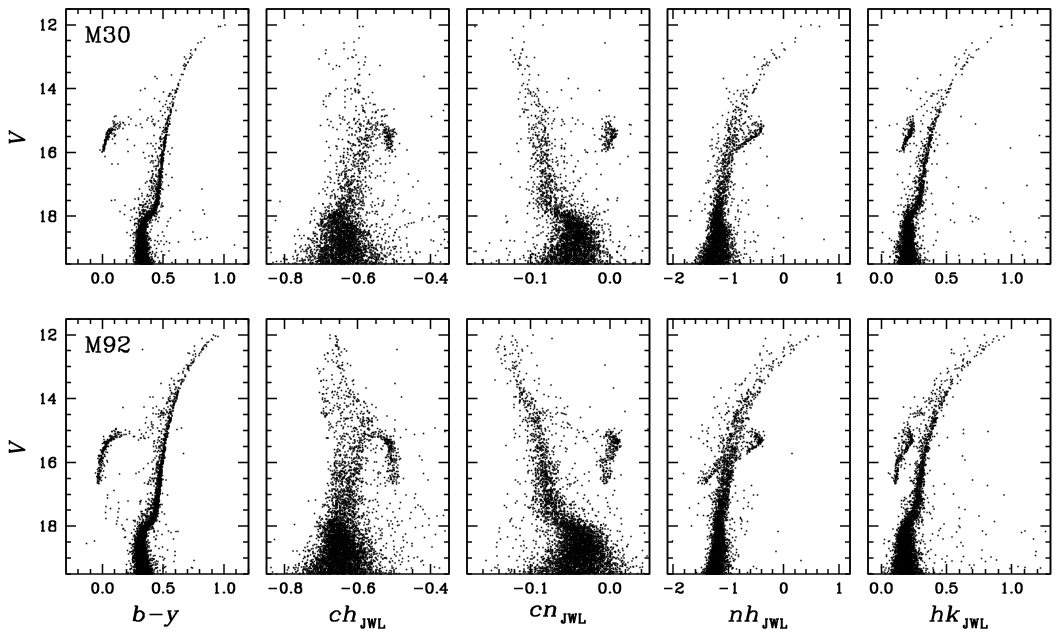}
\caption{(Top) CMDs of M30 proper motion member stars.
(Bottom) CMDs for M92 proper motion member stars.}\label{fig:cmd}
\end{figure*}

In Figure~\ref{fig:by}, we show a comparison of $(b-y)$ color--magnitude diagrams (CMDs) of M30 and M92 by putting them together at the same distance.\footnote{ In the figure, we applied $\Delta(b-y)$ = +0.0084 mag and $\Delta V$ = $-$0.01 mag for M92 with respect to M30 to account the differences in the foreground reddening values and the apparent visual distance moduli between the two clusters \citep{harris96}. Note that \ebv\ = 0.03 and 0.02 for M30 and M92, respectively, and we adopted $E(b-y)$ = 0.84 \ebv\ from our previous study \citep{lee23b}. The apparent visual distance moduli are very similar, 14.64 and 14.65 mag for M30 and M92, respectively.} The figure clearly show the difference in the blue HB (BHB) morphology, while the lack of difference in the red HB and asymptotic giant branch (AGB). M92 and M30 have similar ages and metallicities \citep{sandquist99, dotter10} but different horizontal branch (HB) morphologies, making an exemplary second--parameter pair \citep[e.g.,][]{arp55b, sandage67}. Since mid 1950's, a strong correlation between the HB morphology and the metallicity of the Galactic GCs has been recognized  \citep[e.g.,][]{arp55a}: The metallicity is the first parameter that governs the HB morphology. However, it was soon noticed that the HB morphology of some fraction of Galactic GCs cannot be monotonically parameterized by the metallicity alone, requiring additional parameters. Among several candidates, the age \citep[e.g.,][]{sandage62, sz, ldz94, lee99} and the helium abundance \citep[e.g.,][]{arp55a,  faulkner66, sandage67} have been frequently considered as strong candidates for the second--parameter, although the helium abundance is extremely difficult to measure in the GC stars due to the lack of measurable absorption lines in the cool red giant branch (RGB) stars and metal levitation and helium settling in the hot HB stars \citep[e.g.,][]{grundahl99}. Furthermore, the HB morphology should be understood in the context of the GC MSPs \citep[e.g., see][]{jang15, lee22}.

We will argue that a large amount of the helium enhancement, $\Delta Y$ $\sim$ 0.050, in the SG of the metal-rich groups of stars is responsible for the extended BHB population in M92. We reinterpret the [Eu/Fe] measurements by \citet{kirby23} and find that the metal--poor (MP) and metal--rich (MR) groups of stars in M92 have significantly different [Eu/Fe] abundances with small scatters. This strongly suggests that they are independently formed out of well mixed interstellar media in different environments. In addition to our previous study \citep{lee23a}, our current study may suggest that M92 is not a typical globular cluster but a more complex system, perhaps a merger remnant of two GCs or a remnant nucleus of the metal--poor progenitor dwarf galaxy.

In appendix, we will show comparisons of our photometric \feh\ and \cfe\ of the M92 RGB stars with those from high resolution spectroscopy by \citet{kirby23}, finding that our \fehhk\ and \cfech\ are in excellent correlations with spectroscopic measurements. We  discuss a potential problem with the color--temperature relations by \citet{mucciarelli21}, which predict significantly low effective temperatures for M92. We also address the faulty performance of the $(U-B)-(B-I)$ color index in populational taggings of the metal--poor GCs.

\section{OBSERVATIONS AND DATA REDUCTION}\label{s:reduction}
The journal of observations for M92 is given in \citet{lee23a}.
Observations for M30 were made in 33 nights in 13 runs from 2007 July to 2019 September using the CTIO 1.0 m and KPNO 0.9 m telescopes. Until 2010, we used the filters that were provided by the CTIO. As we noted in our previous work, some of the CTIO's filter were deteriorated due to aging effect \citep[e.g.,][]{lee19b}. We used our own filters since 2011 both in the CTIO and KPNO, and our work presented here will rely on photometric data using our own filters \citep{lee17, lee19a, lee19b, lee21a}. The total integration times for M30 using our own filters were 6680, 13160, 27400, 12500, 9600, 7600 s for \str\ $y$, $b$, \cajwl, JWL39, JWL43, and JWL34, respectively.

The raw data handling was described in detail in our previous works \citep{lee15, lee17}. The photometry of M30 and standard stars were analyzed using \texttt{DAOPHOTII}, \texttt{DAOGROW}, \texttt{ALLSTAR} and \texttt{ALLFRAME}, and the relevant packages \citep{pbs87, pbs94}. The astrometric solutions for individual stars were calculated using the Gaia Data Release 3 \citep[DR3;][]{gaiadr3} and the \texttt{IRAF IMCOORS} package.

In order to select M30 member stars, we made use of the proper-motion study from the Gaia DR3. We derived the mean proper-motion values using iterative sigma-clipping calculations, finding that, in units of milliarcsecond per year, ($\mu_{\rm RA}\times\cos\delta$, $\mu_{\rm decl.}$) = (0.721, $-$7.245) with standard deviations along the major axis of the ellipse of 0.986 mas yr$^{-1}$ and along the minor axis of 0.856 mas yr$^{-1}$. We considered stars within 3$\sigma$ from the mean values to be M30 proper-motion member stars.

In Figure~\ref{fig:cmd}, we compare CMDs of M30 and M92 using our color indices \citep[see][for our color indices]{lee19a, lee21a, lee22}. In the figure, we shows member stars only from the Gaia proper motions. Both clusters have very small interstellar reddening values, and our color indices are insensitive to the variations in the interstellar reddening values \citep{lee23b}, and the differential reddening effects across our science fields are negligibly small. Therefore, the broad RGB widths in the \chjwl, \cnjwl\ and \nhjwl\ are due to the internal variations of the carbon and nitrogen abundances in each GC. Our results suggest that both M30 and M92 have almost identical CMDs in individual color indices except for the extended BHB population of M92. As mentioned above, both clusters have similar metallicities, \feh\ = $-$2.2 and $-$2.3 dex for M30 and M92, respectively \citep{harris96}. The difference in the HB morphology between the two clusters make them a good example of the second--parameter pair, i.e., GCs with different HB morphologies with similar metallicities.

\begin{deluxetable*}{llccccccccccc}
\tablenum{1}
\tablecaption{Photometric data of RGB Stars in M30 and M92} \label{tab:ph}
\tablewidth{0pc}
\tablehead{\colhead{Name} &
\colhead{ID}     & \colhead{$V$}     & \colhead{$(b-y)$} &
\colhead{\cnjwl} & \colhead{\chjwl}  & \colhead{\nhjwl}  &
\colhead{\hkjwl} & \colhead{\fehhk} & \colhead{\cfech}  &
\colhead{\nfenh} & \colhead{R.A.}    & \colhead{decl.} }
\startdata
M30  &  26 &  12.407 &   0.839 &  $-$0.118 &  $-$0.596 &     0.282  &  0.844 &  $-$2.2042 &  $-$0.5351 &   1.4363 &  325.07913208 &  $-$23.18136024  \\
M30  &  29 &  12.546 &   0.813 &  $-$0.103 &  $-$0.610 &     0.275  &  0.816 &  $-$2.1909 &  $-$0.6246 &   1.7438 &  325.07095337 &  $-$23.16836166  \\
M30  &  30 &  12.764 &   0.766 &  $-$0.105 &  $-$0.638 &     0.074  &  0.746 &  $-$2.2058 &  $-$0.7122 &   1.3039 &  325.10092163 &  $-$23.18638992  \\
M30  &  31 &  12.787 &   0.768 &  $-$0.105 &  $-$0.610 &     0.092  &  0.730 &  $-$2.2289 &  $-$0.5553 &   1.5341 &  325.11093140 &  $-$23.24319458  \\
M30  &  35 &  12.947 &   0.738 &  $-$0.115 &  $-$0.605 &  $-$0.095  &  0.673 &  $-$2.2545 &  $-$0.4593 &   1.0460 &  325.04934692 &  $-$23.20969391  \\
M30  &  41 &  13.006 &   0.726 &  $-$0.108 &  $-$0.579 &  $-$0.007  &  0.690 &  $-$2.2154 &  $-$0.3842 &   1.5129 &  325.00881958 &  $-$23.21672249  \\
M30  &  46 &  13.055 &   0.717 &  $-$0.109 &  $-$0.626 &  $-$0.060  &  0.661 &  $-$2.2393 &  $-$0.5789 &   1.3987 &  325.05804443 &  $-$23.15091705  \\
M30  &  50 &  13.160 &   0.706 &  $-$0.117 &  $-$0.580 &  $-$0.235  &  0.655 &  $-$2.2069 &  $-$0.3534 &   0.6801 &  325.08374023 &  $-$23.18958282  \\
M30  &  53 &  13.178 &   0.703 &  $-$0.110 &  $-$0.597 &  $-$0.189  &  0.664 &  $-$2.1794 &  $-$0.4611 &   0.8898 &  325.09738159 &  $-$23.16900063  \\
M30  &  52 &  13.259 &   0.688 &  $-$0.107 &  $-$0.582 &  $-$0.291  &  0.640 &  $-$2.1970 &  $-$0.3546 &   0.5741 &  325.08929443 &  $-$23.17413902  \\
M30  &  59 &  13.306 &   0.681 &  $-$0.097 &  $-$0.650 &  $-$0.327  &  0.629 &  $-$2.1849 &  $-$0.7155 &   0.4486 &  325.09591675 &  $-$23.18000031  \\
M30  &  60 &  13.390 &   0.675 &  $-$0.111 &  $-$0.595 &  $-$0.315  &  0.613 &  $-$2.1971 &  $-$0.3877 &   0.7174 &  325.10140991 &  $-$23.16088867  \\
\enddata
Note - This table is available in its entirely in a machine-readable form in the online journal. A portion is shown here for guidance regarding its form and contents.
\end{deluxetable*}

\begin{figure*}
\epsscale{.9}
\figurenum{3}
\plotone{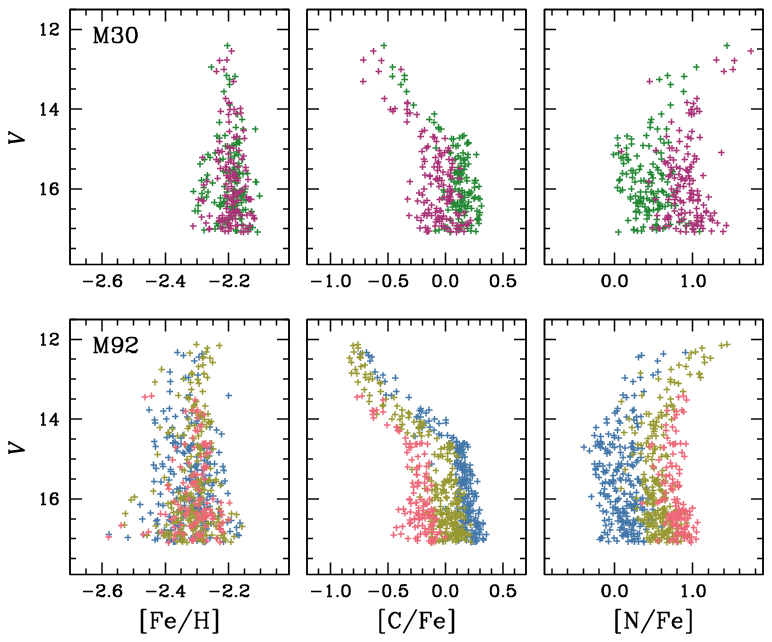}
\caption{(Top) Distributions of photometric elemental abundances against the $V$ magnitude of M30 RGB stars. The green and purple colors denote the FG and SG RGB stars in M30.
(Bottom) Same as top panels, but for M92. The blue, olive, and red colors denote the FG, IG, and SG of the M92 RGB population.}\label{fig:abund}
\end{figure*}

\section{Metallicity, Carbon, and Nitrogen Abundances}
We derive photometric \feh, \cfe, and \nfe\ for M30 using the upgraded method that we applied for M92 \citep{lee23a}. First, we calculate photometric metallicity, \fehhk, of individual RGB stars. We retrieved the model isochrones for [Fe/H] = $-$2.5, $-$2.3, and $-$2.1 dex, $Y$ = 0.247, 0.275, and 0.300 with [$\alpha$/Fe] = +0.4 dex, and the age of 12.5 Gyr from a Bag of Stellar Tracks and Isochrones \citep[BaSTI;][]{basti21}. We adopted different CNO abundances, [C/Fe] = ($-$0.6, $\Delta$[C/Fe] = 0.2 ,0.6), [N/Fe] = ($-$0.8, $\Delta$[N/Fe] = 0.4 , 1.6), and [O/Fe] = (0.1, 0.3, 0.5) for each model grid. We constructed 97 model atmospheres and synthetic spectra for each chemical composition from the lower main sequences to the tip of RGB sequences using \atlas\ \citep{kurucz11}, the latest versions of \moogscat\ \citep{sneden74, moogscat}, and \linemake\ \citep{linemake}. As discussed in our previous studies \citep{lee21a, lee23a}, Rayleigh scattering from the neutral hydrogen (RSNH) atoms in the metal--poor GC RGB stars is the dominant source of the continuum opacity in the blue and ultra-violet wavelength regions. \moogscat\ takes proper care of RSNH with an non-local thermodynamic equilibrium (non-LTE) treatment of the source function. Without the proper care of the RSNH during the synthetic spectrum calculations, some of our color indices, in particular \nhjwl, cannot be reproduced correctly.

In our current study, we upgraded our codes to derive the photometric elemental abundances of individual RGB stars in a more sophisticated way. For photometric metallicity, we assumed the following relation,
\begin{eqnarray}
{\rm [Fe/H]}_{hk} &\approx& f_1({{hk}}_{\mathrm{JWL}},~Y,~\mathrm{[C,N,O/Fe]},~M_V).\label{eq:FeH}
\end{eqnarray}
This relation makes use of more input parameters than that used in our previous study \citep{lee21a},
\begin{eqnarray}
{\rm [Fe/H]}_{hk} &\approx& f_1^\prime({{hk}}_{\mathrm{JWL}},~M_V).\label{eq:oldFeH}
\end{eqnarray}
At first, the CNO and helium abundances are assumed to be the same for all RGB stars. Then the carbon and nitrogen abundances are updated using the \cfech\ and \nfenh\ from the relations given below and the helium abundances estimated from their RGB bump (RGBB) $V$ magnitudes. Since we do not have oxygen abundances of individual stars, we assigned [O/Fe] = 0.5 dex for the FG and 0.0 dex for the SG. We note that the presumed oxygen abundance does not significantly affect our results.
We iterated these processes until the derived metallicity converges within $\leq$ 0.01 dex. In most cases, no more than 3 iterations is necessary. The difference in the \fehhk\ value from our current study for M92 and that of our previous study \citep{lee23a} is \dfeh$_{\rm hk}$ = $-$0.005 $\pm$ 0.002 dex ($\sigma$ = 0.057), in the sense of the current minus the previous work.

We obtained \fehhk\ = $-$2.194$\pm$0.002 dex ($\sigma$ = 0.041) for M30. Our metallicity for M30 is about 0.1 dex larger than that of M92 (see Table~\ref{tab:ab}), $-$2.316$\pm$0.003 dex ($\sigma$ = 0.062). Unlike M92, M30 appears to have an unimodal metallicity distribution as shown in Figure~\ref{fig:abund}. Note that the metallicity dispersion of M30 is significantly smaller than that of M92. It is similar to those of two metallicity groups (metal--poor:MP, metal--rich:MR) in M92, $\sigma$ = 0.044 and 0.042 dex \citep[see Table~1 of][]{lee23a}.

\begin{deluxetable*}{lrrr}
\tablenum{2}
\tablecaption{
Comparisons of Elemental Abundances of RGB Stars in M30 and M92} \label{tab:ab}
\tablewidth{0pc}
\tablehead{\colhead{} & \colhead{\fehhk} & \colhead{\cfech\tablenotemark{a}} & \colhead{\nfenh\tablenotemark{a}}
}
\startdata
M30 & $-$2.194$\pm$0.002($\pm$0.041) & 0.053$\pm$0.009($\pm$0.134) & 0.657$\pm$0.021($\pm$0.325) \\
M92 &  $-$2.316$\pm$0.003($\pm$0.062) & 0.031$\pm$0.008($\pm$0.169) & 0.443$\pm$0.015($\pm$0.323) \\
M92-MP &  $-$2.414$\pm$0.004($\pm$0.044) & 0.072$\pm$0.019($\pm$0.173) & 0.404$\pm$0.038($\pm$0.343) \\
M92-MR &  $-$2.295$\pm$0.002($\pm$0.042) & 0.023$\pm$0.009($\pm$0.167) & 0.452$\pm$0.016($\pm$0.318) \\
\enddata
\tablenotetext{a}{For RGB stars fainter than their RGBB $V$ magnitudes.}
\end{deluxetable*}

The photometric carbon and nitrogen abundances for M30 RGB stars were estimated using the following upgraded relations,
\begin{eqnarray}
{\rm [C/Fe]}_{ch} &\approx& f_2({{ch}}_{\mathrm{JWL}},~{\mathrm{[Fe/H]}_{hk}},~Y,~\mathrm{[N,O/Fe]}, ~ M_V), \\
{\rm [N/Fe]}_{nh} &\approx& f_3({{nh}}_{\mathrm{JWL}},~{\mathrm{[Fe/H]}_{hk}},~Y,~\mathrm{[C,O/Fe]}, ~ M_V).
\end{eqnarray}
The \cfech\ and \nfenh\ returned from our current relations provide slightly different results than those from our previous study. For M92, the differences are \dcfe\ = 0.016 $\pm$ 0.007 dex ($\sigma$ = 0.181) and \dnfe\ = 0.017 $\pm$ 0.009 dex ($\sigma$ = 0.226).
In Table~\ref{tab:ph}, we show our photometric elemental abundances along with our photometry for RGB stars with \vvhbmag\ in M30 and M92.
In Figure~\ref{fig:abund}, the evolution of the carbon and nitrogen abundances against the $V$ magnitude can be clearly seen due to the deep mixing accompanied by the CN hydrogen shell burning. In Table~\ref{tab:ab}, we show the mean and dispersion of \cfech\ and \nfenh\ for the two clusters. The mean \cfech\ values of the RGB stars fainter than their RGBB $V$ magnitudes are similar for both clusters, suggesting that their primordial mean carbon abundances are the same. However, the \cfe\ dispersion of M92 is slightly larger than that of M30, which reflects the fact that M92 has a more extremely enhanced population as shown below. On the other hand, the mean primordial nitrogen abundance of the RGB stars in M30 is \dnfe\ $\sim$ 0.2 dex larger than those in M92, indicating that M30 and M92 formed out of interstellar media experienced different chemical enrichment histories.

In Appendix~\ref{ap:m92}, we show comparisons of our \feh\ and \cfe\ of M92 RGB stars with those from high resolution spectroscopy by \citet{kirby23}.
We obtained that our \fehhk\ is nicely correlated with their \feh, finding $\langle$\fehhk\ $-$ \feh$_{\rm K23}\rangle$ = 0.183$\pm$0.007 dex ($\sigma$ = 0.031 dex, Figure~\ref{fig:kirby}(g)--(i)), and a Pearson correlation coefficient of 0.825. We note the small dispersion in the mean difference, a strong indication of the reliability of our measurements. The rather large offset between the two studies is partially due to inappropriate effective temperatures used by \citet{kirby23} as discussed in Appendix B.

For our carbon abundance measurements, we achieved an excellent agreement with those of \citet{kirby23}, finding that $\langle$\cfech\ $-$ \cfe$_{\rm K23}\rangle$ = $-0.103\pm0.016$ dex ($\sigma$ = 0.073 dex, Figure~\ref{fig:kirby}(j)--(l)), with a Pearson correlation coefficient of 0.981. Again, it is believed that the inappropriate effective temperature adopted by \citep{kirby23} is responsible for the offset between the two measurements, at least in some part. Again, we stress that the scatter in the $\langle$\cfech\ $-$ \cfe$_{\rm K23}\rangle$ around the mean value is very small. Therefore, it should be fair to say that our photometric carbon abundances is very accurate in a relative sense, at least.

\begin{figure*}
\epsscale{1.1}
\figurenum{4}
\plotone{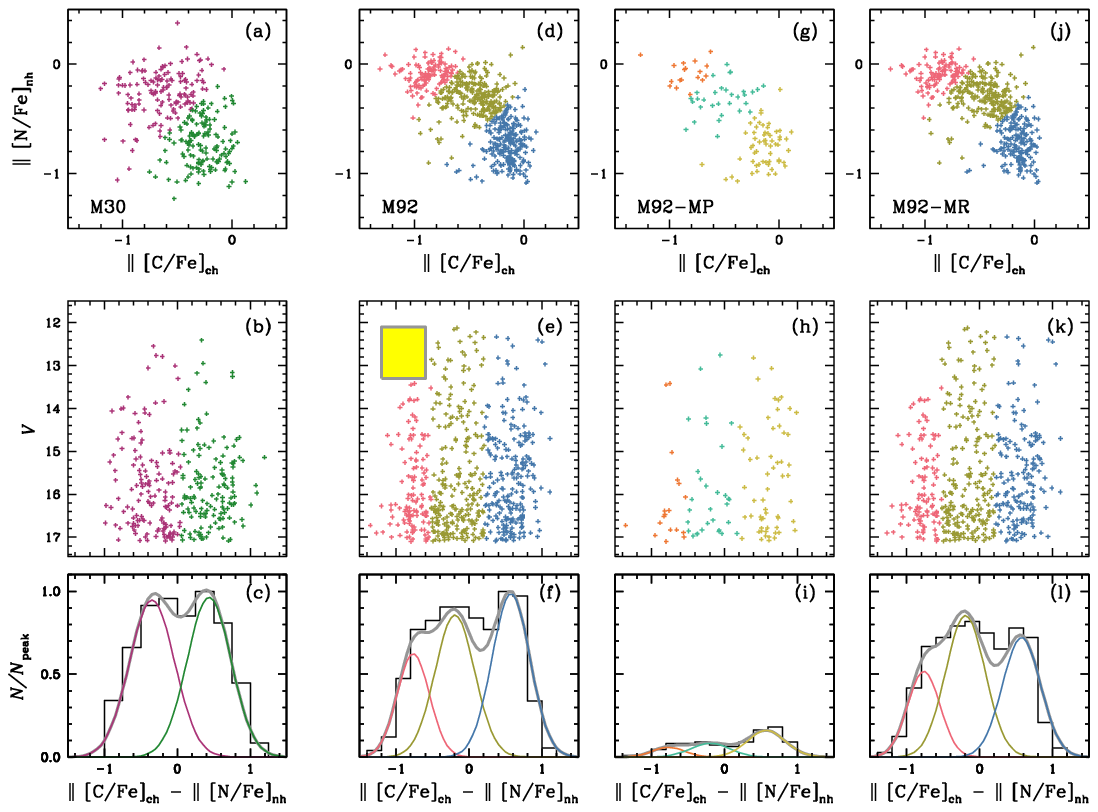}
\caption{(a) A plot of \pcfe\ vs. \pnfe\ of the M30 RGB stars with \vvhbmag. The green and purple colors denote the FG and SG RGB stars in M30. (b) A plot of \dcfenfe\ vs. $V$ of the M30 RGB stars. (c) A histogram for the \dcfenfe\ distribution. The green and purple solid lines denote those returned from our EM estimator, while the gray solid line is for the total RGB population.
(d)--(f) Same as (a)--(c), but for M92. The blue, olive, and red colors denote the FG, IG, and SG of the M92 RGB population. (e) The bright yellow shaded box indicates the region where no RGB stars present, due to high helium abundance of the M92 SG.
(g)--(i) Same as (a)--(c), but for the M92 MP population. The yellow, mint, and orange colors denote the FG, IG, and SG of the M92 MP RGB population.
(j)--(l) Same as (a)--(c), but for the M92 MR population. The blue, olive, and red colors denote the FG, IG, and SG of the M92 MR RGB population.}\label{fig:em}
\end{figure*}

\section{Populational Taggings}\label{sec:tagging}
Due to internal mixing episodes accompanied by the CN cycle in the hydrogen shell burning region, the surface carbon and nitrogen abundances of the RGB stars brighter than their RGBBs are significantly changed from their initial abundances, as Figure~\ref{fig:abund} represents well. Furthermore, the mixing efficiency sensitively depends on metallicity since the hydrogen shell will be hotter and wider with decreasing metallicity \citep[e.g., see][and references therein]{lee23b}. Therefore, it is difficult to classify individual populations for the bright RGB stars from their \cfe\ and \nfe\ as already shown in Figure~\ref{fig:abund}. Similar to our previous approach, we make use of the normalized abundance ratios \pcfe\ and \pnfe. We derived the outer envelopes of our \cfech\ and \nfenh\ against $V$ magnitude bins and they were fitted with the second order polynomials for those of fainter and brighter than the RGBB \citep[see also][]{milone17}. Then the normalized \cfech\ and \nfenh\ are calculated as
\begin{eqnarray}
\parallel {\rm [C/Fe]}_{\rm ch} &\equiv& \frac{{\rm [C/Fe]}_{\rm ch} - {\rm [C/Fe]}_{\rm ch,max}} {{\rm [C/Fe]}_{\rm ch, max}-{\rm [C/Fe]}_{\rm ch, min}},\label{eq:pcfe}\\
\parallel {\rm [N/Fe]}_{\rm nh} &\equiv& \frac{{\rm [N/Fe]}_{\rm nh} - {\rm [N/Fe]}_{\rm nh,max}} {{\rm [N/Fe]}_{\rm nh, max}-{\rm [N/Fe]}_{\rm nh, min}},\label{eq:pnfe}
\end{eqnarray}
where the subscripts denote the fiducials of the maximum and the minimum sequences of the \cfech\ and \nfenh.

In Figure~\ref{fig:em}, we show plots of the \pcfe\ versus \pnfe\ and the \dcfenfe\ versus $V$ magnitude for both clusters. In Figure~\ref{fig:em}(e), we marked a yellow box where no bright RGB stars are present. As we will show later, this yellow box region is supposed to be the region where bright RGB stars with enhanced helium abundance (the M92--SG) by \dy\ $\sim$ 0.05 would pass through.
However, the helium enhanced stars have a denser core with a higher central temperature due to a larger mean molecular weight. As a consequence, the helium flash occurs at a smaller core mass, resulting in fainter RGB tip luminosity as the helium abundance increases \citep[e.g., see][]{valcarce12}. We believe that the lack of bright RGB stars in the yellow box is a strong evidence of a large helium enhancement in M92.

Finally, by generating histograms for the (\dcfenfe), we applied an expectation--maximization (EM) algorithm for two--component Gaussian mixture models for the M30, while we applied a three--component Gaussian mixture model for the M92 \citep[e.g., see][]{lee23a}. Through these calculations, we obtained populational number ratios of \nrgb\ = 50.2:49.8 ($\pm$5.1) for M30. Our result is in good agreement with that by \citet{carretta09}, who obtained \nrgb\ = 41($\pm$12):58($\pm$14) from the [Na/Fe] and [O/Fe] measurements, and \citet{leitinger23}, who obtained \nrgb\ = 46:54($\pm$5) from a combined photometric of \citet{pbs19} and \citet{nardiello18}. In Appendix~C, we discuss the problem of applying the \cubi\ index for the metal-poor GCs. As shown in Figure~\ref{fig:emcubi}, the populational number ratio by \citet{leitinger23} could be a coincidence. The critical factor that we have to pursue is the proper population assignment of the individual stars, not the hazy overall populational number ratios.
On the other, our populational ratio is slightly different from that by \citet{milone17}, 38.0:62.0 ($\pm$2.8), and is significantly different from that derived from Washington photometry by \citet{frelijj16}, \nrgb\ $\sim$ 15:85. It is not surprising because, Washington photometry is not suitable to study GC MPs, in particular for those in metal--poor regimes same as $UBI$ photometry, as we discussed in Appendix~\ref{ap:cubi}.
We also note that \citet{carretta09} classified 3$\pm$3 extreme component RGB stars out of 99 stars in M30, but our \dcfenfe\ histogram does not clearly indicate such population.

For M92, we obtained a \nthree\footnote{Our FG, IG, and SG are approximately equivalent to the primordial, intermediate, and extreme populations devised by \citet{carretta09}. See also \citet{lee21b}.} = 38.1:35.5:26.4 ($\pm$2.7), where the IG denotes the intermediate generation of stars. Note that our current estimation for  M92 is slightly different from our previous value from the RGB stars fainter than the RGBB, 32.2:31.6:36.2 ($\pm$2.4) \citep{lee23a}, for two reasons. First, in our current study, we upgraded our computer programs to calculate photometric abundances in a more sophisticated way and our current photometric elemental abundances for M92 RGB stars are slightly different from those in our previous study \citep{lee23a}. Second, in our previous study, we relied on the (\cfech\ $-$ \nfenh), not the (\dcfenfe), since we were interested in the RGB stars fainter than the RGBB and the normalization processes such as in Equations~\ref{eq:pcfe} and \ref{eq:pnfe} were not applied in our previous study.

Our populational number ratio here is different from that of \citet{milone17}, \nrgb\ = 30.4:69.6($\pm$1.5), who decomposed the FG and SG only assuming that M92 is a mono-metallic GC, and that of \citet{leitinger23}, who obtained \nrgb\ = 45:55($\pm$3) for M92.

We note that the populational number ratios of the MP and MR populations in M92 are significantly different: \nthree\ = 49:28:23 for the MP and 36:37:27 for the MR. As we pointed out in our previous work \citep{lee23a}, given the current mass of M92 \citep{baumgardt18}, the total mass of the M92--MP population is $\sim$ 6$\times$10$^{4}$\msun, comparable to those of less massive Galactic GCs with the FG population dominance \citep[e.g., see][]{milone20}.

\begin{figure*}
\epsscale{.9}
\figurenum{5}
\plotone{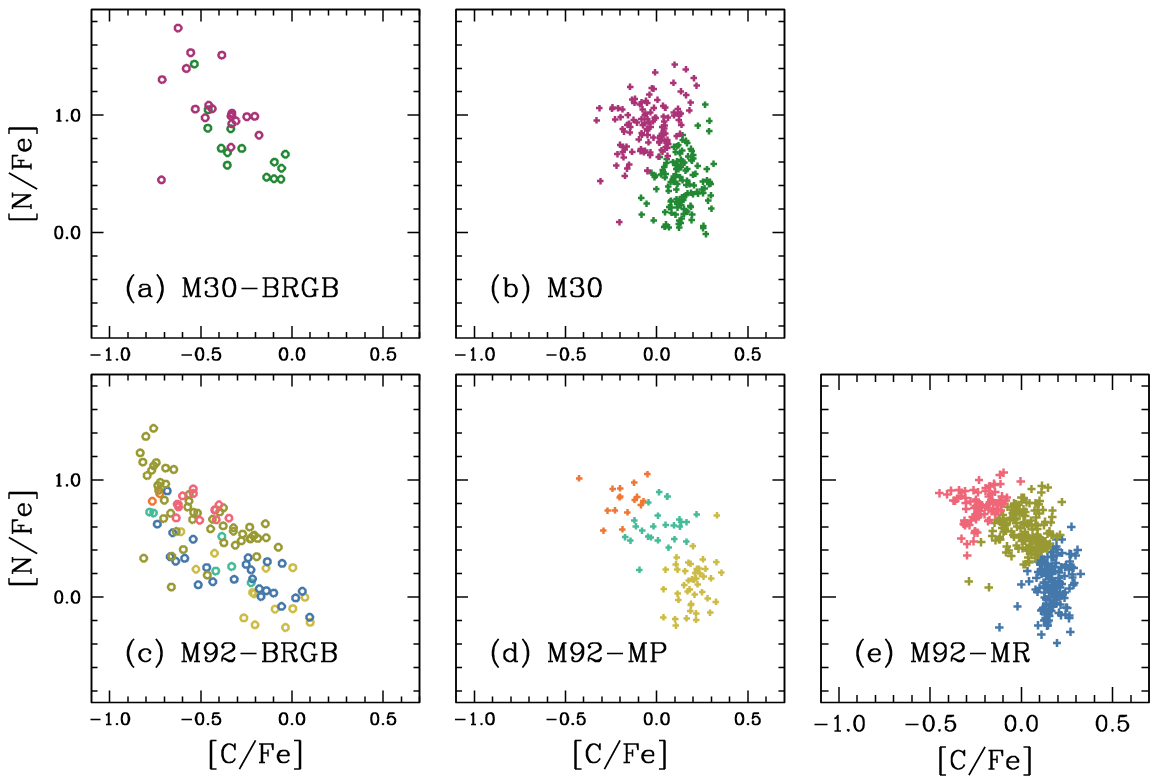}
\caption{
Plots of [C/Fe] vs. [N/Fe] of M30 and M92.
(a) A plot of \cfech\ vs. \nfenh\ for bright RGB stars (i.e., those brighter than the RGBB) in M30 (green:FG, purple:SG).
(b) Same as (a) but for the faint RGB stars (i.e., those fainter than the RGBB) in M30.
(c) Same as (a) but for M92  (yellow:MP-FG, mint:MP-SG, orange:MP-SG, blue:MR-FG, olive:MR-IG, red:MR-SG).
(d) Same as (a) but for the faint RGB stars in the M92--MP.
(e) Same as (a) but for the faint RGB stars in the M92--MR.
}\label{fig:cfenfe}
\end{figure*}

In Figure~\ref{fig:cfenfe}, we show plots of [C/Fe] versus [N/Fe] of individual populations in M30 and M92. It is clear that M92 exhibits a more extended C--N anticorrelation than M30 does, which is related to the extended BHB population in M92.

\begin{figure*}
\epsscale{1.1}
\figurenum{6}
\plotone{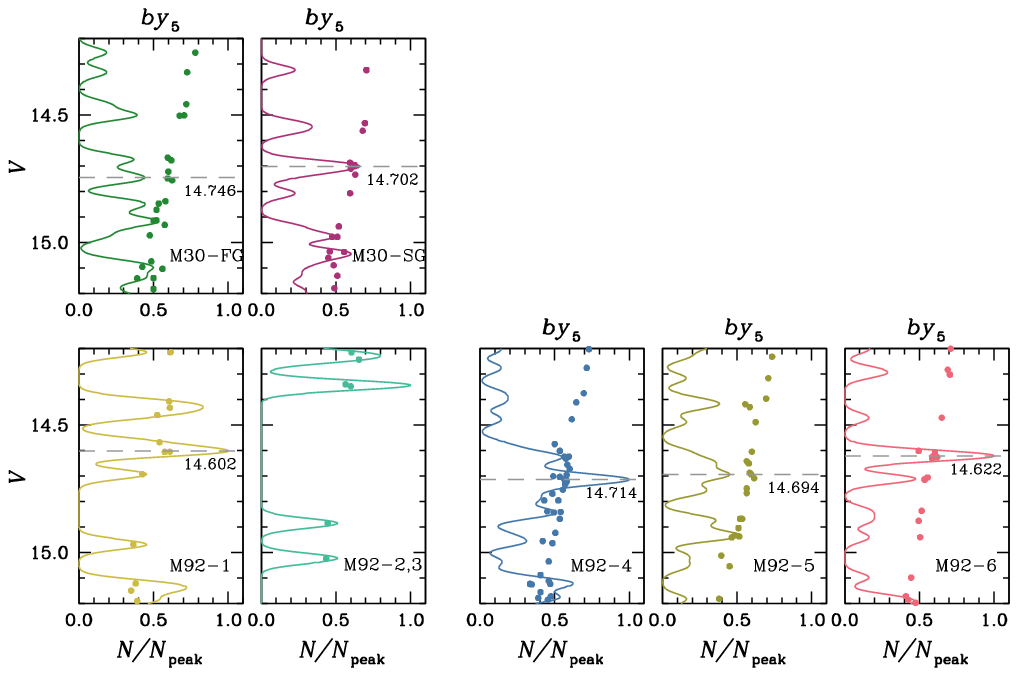}
\caption{
(Top) Plots of the $by5$ vs. $V$ CMDs and smoothed differential LFs for the M30--FG and SG. The gray dashed lines indicate the RGBB $V$ magnitudes of individual populations.
(Bottom) Same as the top panels but for M92 (M92--1:MP--FG, 2:MP--IG, 3:MP--SG, 4:MR--FG, 5:MR--IG, and 6:MR--SG).
}\label{fig:rgbb}
\end{figure*}

\begin{figure}
\epsscale{1.2}
\figurenum{7}
\plotone{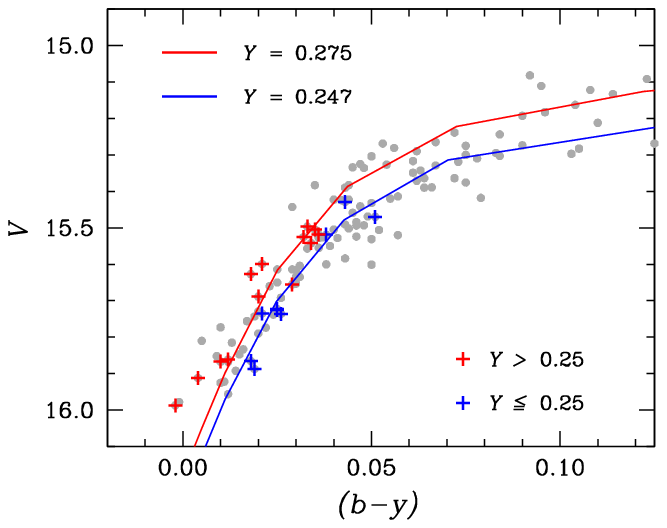}
\caption{The CMD of the M30 HB region. We show HB stars with helium abundances \citep{mucciarelli14}. Thin solid lines are for the ZAHB loci for $Y$ = 0.247 and 0.275 \citep{basti21}. Note that we exclude the star 30016296,  which has a very low helium abundance, $Y$ = 0.088$\pm$0.022.
}\label{fig:m30}
\end{figure}

\section{Red-Giant Branch Bump}
During the course of low--mass star evolution, RGB stars experience a temporary drop in luminosity when the very thin hydrogen burning shell crosses the discontinuity in the chemical composition and lowered mean molecular weight left by the deepest penetration of the convective envelope during the ascent of the RGB.
In such evolutionary phases, RGB stars have to cross three times the same luminosity interval, leaving a distinctive feature, the so-called RGBB \citep[e.g., see][]{cassisi13}. It is well recognized that, at a given age, the RGBB luminosity increases with helium abundance and decreases with metallicity \citep[e.g., see][]{cassisi97}.

The prominence of the RGBB also depends on metallicity, in the sense that the RGBB bump feature becomes more distinguishable with increasing metallicity \citep[e.g., see][]{fusipecci90, valcarce12}. At the same time, the number of RGB stars in M30 is much smaller than those in M92, for example, as shown in Figure~\ref{fig:cmd}.\footnote{The current mass of M30 is about a half of that of M92 \citep{baumgardt18}. The numbers of RGB stars used by Milone et al. (2017) and Leithinger et al.\ (2023) for their studies also reflect the mass difference between the two clusters: $n$(M92):$n$(M30) = 795:323 \citep{milone17} and  740:295 for the HST and 238:110 for the ground-based catalogues that \citet{leitinger23} used. Note that we used $n$(M92):$n$(M30) = 592:271.} We suspect that, due to these reasons, \citet{sandquist99} did not detect clear signs of the RGBB in M30.

In Figure~\ref{fig:rgbb}, we show the CMDs and smoothed differential luminosity functions (LFs) with Gaussian kernels around the RGBB regimes of individual populations in M30 and M92. In the figure, we used a modified color $by5$ (= $5\times(b-y)-2.2$) for the clarity of the RGB distributions.
In order to interpret the observed RGBBs, we relied on Monte Carlo simulations by constructing evolutionary population synthesis models \citep[e.g.,][]{lee22}.
We retrieved BaSTI isochrones for \feh\ = $-$2.5, $-$2.3, and $-$2.1 and $Y$ = 0.247, 0.275, and 0.300 with the age of 12.5 Gyr \citep[e.g.,][]{dotter10}. To determine the \vbump\ of individual model isochrones, we populated 200,000 artificial stars and we generated generalized histograms to derive the \vbump\ as we did for our observed data. We obtained \dvbump/\dfeh\ $\propto$ 0.560 $\pm$ 0.002 and \dvbump/\dy\ $\propto$ $-$1.786 $\pm$ 0.002. Using these relations, we calculate the RGBB $V$ magnitude differences in terms of relative metallicity and helium abundance. We summarize our results as follow:
\begin{itemize}\setlength\itemsep{0em}
\item The RGBB $V$ magnitude difference \dvbump\ = 0.032$\pm$0.028 mag\footnote{The apparent distance modulus of M92 is 0.01 mag larger than that of M30 \citep{harris96}}  between the M30 FG and M92 MR--FG can be interpreted by the metallicity difference between the two populations, \dfeh$_{\rm bump}$ = 0.057$\pm$0.051 dex, assuming they have the identical helium abundance, $Y$ = 0.247. The metallicity difference between M30 FG and M92 MR using the RGBB $V$ magnitude is in agreement with that from our photometric metallicity, $\Delta$\fehhk\ = 0.101$\pm$0.004 dex.

\item \dvbump\ = 0.044 mag between the M30 FG and SG can be explained by the helium abundance difference of \dy$_{\rm bump}$ = 0.025$\pm$0.016, which is in excellent agreement with that of \citet{milone18}, $\delta Y_{\rm max}$ = 0.022$\pm$0.010.

\item \dvbump\ = 0.114 mag between the M92 MP and MR--FG can be explained by the metallicity difference of \dfeh$_{\rm bump}$ = 0.200$\pm$0.051 dex, assuming they have the identical helium abundance, $Y$ = 0.247. In our previous study \citep{lee23a}, we argued that M92 exhibits a bimodal metallicity distribution with $\langle$\dfeh$\rangle$ = 0.119$\pm$0.004 dex ($\sigma$ = 0.061), which is slightly smaller than the current estimate using the RGBB $V$ magnitude difference between the M92 MP and MR--FG.

\item \dvbump\ = 0.020 and 0.092 mag between the M92 MR--FG and IG, SG can be translated by the helium abundance difference of \dy$_{\rm bump}$ = 0.011 and 0.052$\pm$0.016, which is marginally in agreement with those of \citet{milone18}, \dy$_{\rm max}$ = 0.039$\pm$0.016 (color spreads in the RGB), \citet{vandenberg22}, $\Delta Y$ $\approx$ 0.04 (isochrone fittings), and \citet{ziliotto23}, \dy\ $\approx$ 0.01--0.04 (color spreads in the lower main--sequence).
\end{itemize}

\begin{figure}
\epsscale{1.2}
\figurenum{8}
\plotone{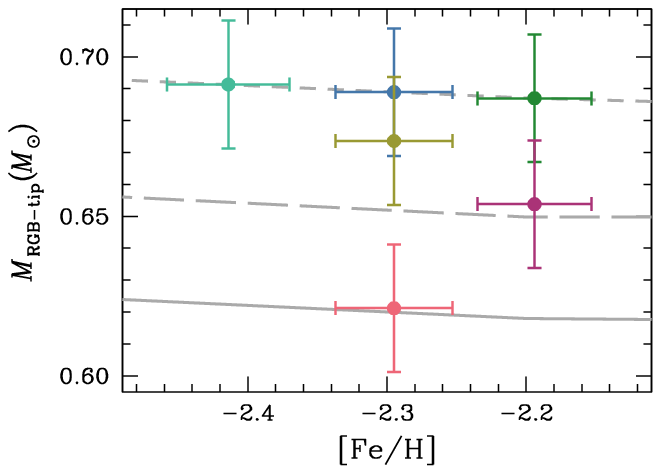}
\caption{
RGB tip masses against [Fe/H]. The short and long dashed, and solid lines denote for $Y$ = 0.247, 0.275, and 0.300, respectively. The colors are the same as Figure~\ref{fig:cfenfe}. }\label{fig:tip}
\end{figure}

\begin{deluxetable*}{llccccccc}
\tablenum{3}
\tablecaption{
Input Parameters Used in Our Synthetic HB Model Calculations} \label{tab:synhb}
\tablewidth{0pc}
\tablehead{\colhead{Cluster} & \colhead{Pop.} & \colhead{\feh} &  \colhead{$\sigma$\feh} &  \colhead{$Y$} & \colhead{$\sigma Y$} & \colhead{$\langle M_\odot \rangle$} & \colhead{$\sigma M_\odot$} & \colhead {frac.(\%)}
}
\startdata
M30 & FG    & $-$2.19 & 0.044 & 0.247 & 0.0025 & 0.687 & 0.020 & 50 \\
M30 & SG    & $-$2.19 & 0.038 & 0.272 & 0.0025 & 0.654 & 0.020 & 50 \\
\hline
M92 & MP     & $-$2.41 & 0.044 & 0.247 & 0.0025 & 0.691 & 0.020 & 19 \\
M92 & MR--FG & $-$2.29 & 0.042 & 0.247 & 0.0025 & 0.689 & 0.021 & 32 \\
M92 & MR--IG & $-$2.29 & 0.042 & 0.258 & 0.0025 & 0.674 & 0.020 & 31 \\
M92 & MR--SG & $-$2.29 & 0.042 & 0.299 & 0.0025 & 0.621 & 0.018 & 18 \\
\enddata
\end{deluxetable*}

\begin{figure}
\epsscale{1.2}
\figurenum{9}
\plotone{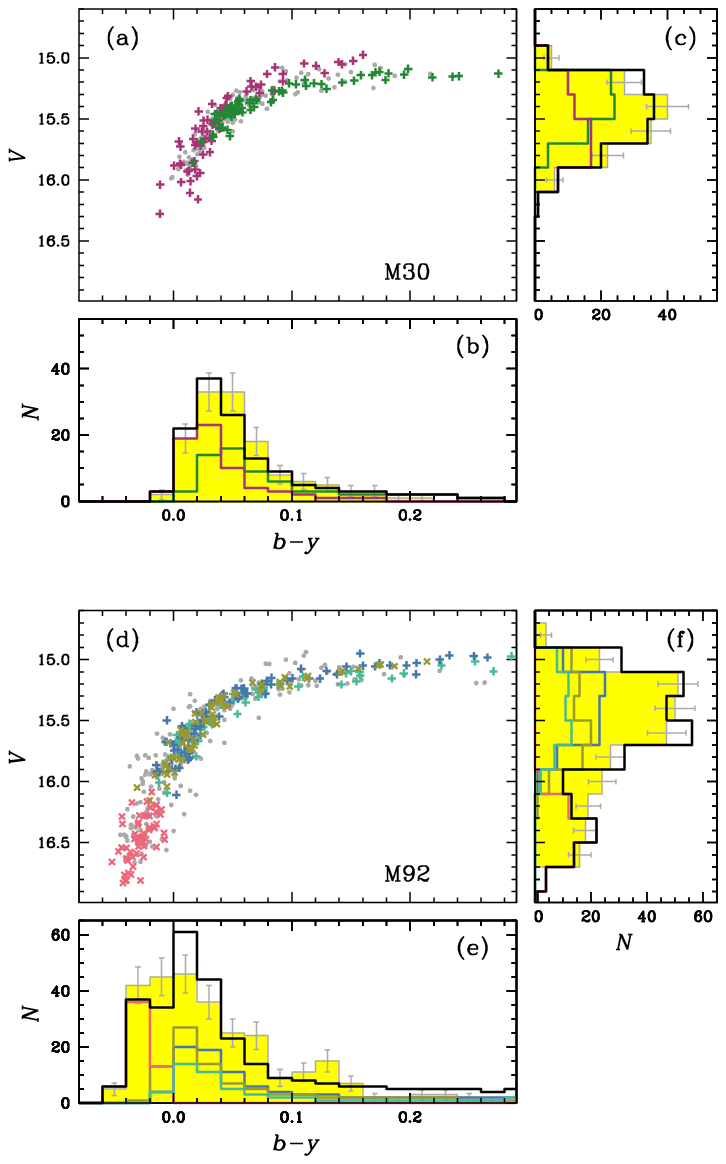}
\caption{
(a) A comparison of observed (gray dots) and synthetic (the same color scheme as Figure~\ref{fig:cfenfe}) CMDs of M30 HB stars.
(b) The yellow shaded histogram with error bars denotes the observed ($b-y$) distribution of M30 HB stars, while the green, purple, and black histograms denote the  mean histograms of 100 trials for the synthetic FG, SG, and all HB stars.
(c) Same as (b) but for the $V$ magnitude.
(d) Same as (a) but for M92. The yellow, mint, blue, olive, and red plus signs denote the synthetic MP--FG, MP--SG, MR--FG, MR--IG, and MR--SG HB stars.
(e) Same as (b), but for M92.
(f) Same as (c), but for M92.
}\label{fig:synhb}
\end{figure}

\section{Horizontal Branch}
HB stars can play a significantly role to probe the metallicity and helium abundance in GCs, due to their rather sensitive morphological dependence on chemical compositions. Figure~\ref{fig:m30} shows the CMD of the M30 HB region along with BHB stars with helium abundance measurements from intermediate-resolution ($R \sim$ 18,000) spectroscopy by \citet{mucciarelli14}. In the figure, we also show the zero-age HB (ZAHB) loci for $Y$ = 0.247 and 0.275 \citep{basti21}. The figure clearly shows that the M30 BHB stars are segregated by their helium abundances, in the sense that the BHB stars with lower helium abundances occupy the redder part of the BHB domain at a given $V$ magnitude, consistent with the model prediction. This also suggests that our estimates of the helium abundance dispersion from the RGBB $V$ magnitudes are fairly reasonable.

Encouraged with Figure~\ref{fig:m30}, we constructed synthetic HB models following the similar methods that we developed in our previous studies \citep[e.g.,][]{lee99, lee22}, in order to interpret the observed HB morphologies of M30 and M92.

First, we obtained the RGB tip masses for individual populations using the BaSTI isochrones with the Reimer's index of $\eta$ = 0.3 and [$\alpha$/Fe] = +0.4. For this purpose, we adopt relative helium abundances returned from our RGBB $V$ magnitudes. In Figure~\ref{fig:tip}, we show the RGB tip masses with different helium abundances as functions of metallicity. In the figure, we also show the mean masses of individual populations in M30 and M92 (the error bars indicate $\pm\sigma$). Next, we populate the synthetic HB stars with our evolutionary population synthesis models using the input parameters as shown in Table~\ref{tab:synhb}.

In Figure~\ref{fig:synhb}, we show our results for synthetic HB models along with the observed HB stars in both clusters. In the figure, our synthetic histograms are the mean values returned from 100 trials, while the synthetic CMDs are those of one specific trial. We note that our simulations reproduce the observed HB distributions of both clusters reasonably well. We also calculated the synthetic HB models for M92 using a mildly enhanced helium abundance for the MR--SG, \dy\ = 0.02--0.03, which fail to reproduce the observed extended BHB distribution with $V$ $\gtrsim$ 16.1 mag in M92.\footnote{Note that, at [Fe/H] $\sim$ $-$2.20 dex and $Y$ = 0.300, the HB stars less massive than $\sim$ 0.510 \msun\ fail to evolve into the normal AGB phases and become AGB-manqu\'e stars. Their masses are significantly smaller than the HB masses presented in Table~\ref{tab:synhb}, and the emergence of the AGB-manqu\'e does not affect our simulations.}

The fraction of the underlying populations may also suggest that the extended BHB is related to the helium enhanced RGB population. The fraction of the M92 HB stars fainter than $V$ = 16.1 mag is about 0.20($\pm$0.03), consistent with the fraction of M92--MR--SG, 0.18($\pm$0.02).

We conclude that not only the RGBB $V$ magnitude of the M92 MR--SG but also the presence of the extended BHB population with $V$ $\gtrsim$ 16.1 mag strongly suggest that M92 contains an extreme enhanced population with \dy\ $\approx$ 0.05.

\begin{figure*}
\epsscale{1.1}
\figurenum{10}
\plotone{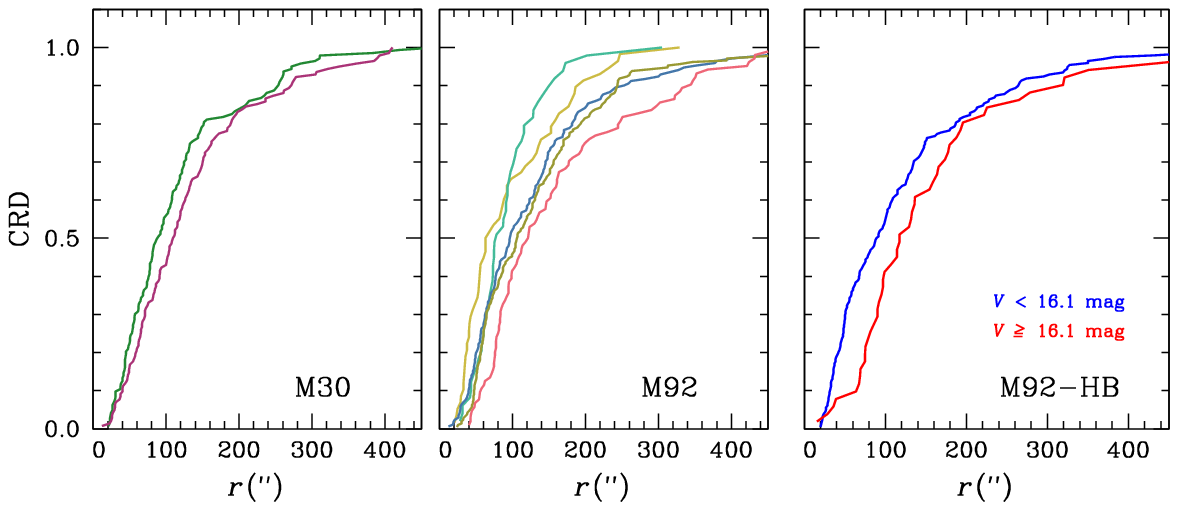}
\caption{
CRDs of individual populations in M30 and M92 RGB stars, and M92 HB stars.
Colors for RGB stars are the same as Figure~\ref{fig:cfenfe}.}\label{fig:crd}
\end{figure*}

\section{Cumulative Radial Distributions}
The observed cumulative radial distributions (CRDs) of individual GC MSPs may depend on many physical properties at the time of GC formation and provide critical information on the dynamical evolution of GCs. Previous studies on the formation of the MSPs in GCs suggested that the SG populations may form in the innermost part of the cluster in a more extended FG system \citep[e.g., see][]{dercole08, bekki19}, where the degree of the helium enhancement of the SG can be dependent on the external gas density \citep{calura19}. The initial structural difference between the FG and SG populations can be gradually erased with time, due to the result of the preferential loss of the FG stars during the cluster's dynamical evolution \citep[e.g., see][]{vesperini21}. It is also likely that the current location can be affected by the different evolutionary stellar masses of individual populations due to different degree of diffusion processes. Over the Hubble time, for example, the radial distributions of the later generations of stars with enhanced helium abundances can expand outward due to the smaller stellar masses resulted from their fast stellar evolution \citep[e.g., see][]{fare18}.

In this regards, we derive the CRDs of individual MSPs, and we show our results in Figure~\ref{fig:crd} and Table~\ref{tab:ks}. The figure shows that the M30 FG is distributed more spatially concentrated than the M30 SG is. However, it does not appear to be statistically significant. We performed the Kolmogorov--Smirnov (K--S) tests and we obtained the $p$ value of $p$ = 0.127, indicating that they may be drawn from the same parent distribution.

The CRDs for the M92 MSPs are more interesting. The M92--MP is more centrally concentrated than the M92--MR is, with the $p$ value of $p$ = 1.98$\times10^{-4}$, strongly suggesting that they are statistically independent groups of stars. This is  apparently consistent with the idea that M92 may be a merger remnant of the two former dwarf galaxy GCs with slightly different formation epoch and structural parameters \citep{lee23a}. Among the M92--MR populations, the FG is more centrally concentrated than the SG, with a $p$ = 0.003, which appears to be consistent with the recent idea proposed by \citet{fare18} that the secular evolution of the CRDs in GC MSPs can be governed by their stellar masses with different helium contents. The small current stellar masses of the M92--MR--SG due to a significant enhanced helium abundance (\dy\ $\sim$ 0.05) is thought to be responsible for its more centrally extended distribution than other populations (also see Table~\ref{tab:synhb}).

We note that \citet{leitinger23}  claimed that the M92 SG is more centrally concentrated than the FG is, in sharp contrast to our result. Again, we believe that this vividly shows the reason for not using broadband photometry for the metal--poor GC MSPs as discussed in Appendix~\ref{ap:cubi}.

We also compared the CRDs of the M92 HB stars. As before, we divide the HB stars with $V$ = 16.1 mag; the bright HB and faint HB. The faint HB is more centrally extended than the bright HB is, with a $p$ value of 6.79$\times10^{-4}$ returned from the K--S test, strongly suggest that they are drawn from different parent distributions. Our K--S test indicates that CRDs of M92--MR--SG and the faint HB are very similar with the $p$ value of 0.980. Our investigation of CRDs clearly tells that the faint HB population is progeny of the helium enhanced population M92--MR--SG.

\begin{deluxetable}{lllc}
\tablenum{4}
\tablecaption{
$p$ Values Returned from the K-S Tests for the CRDs of Individual Populations} \label{tab:ks}
\tablewidth{0pc}
\tablehead{\colhead{Cluster} & \colhead{Pop. 1} &  \colhead{Pop. 2} &  \colhead{$p$ Value} }
\startdata
M30 & FG    &  SG & 0.127 \\
\hline
M92 & MP        &  MR & 1.98$\times10^{-4}$ \\
M92 & MP--FG    &  MP--SG & 0.064 \\
M92 & MR--FG    &  MR--IG & 0.447 \\
M92 & MR--FG    &  MR--SG & 0.003 \\
M92 & MR--IG    &  MR--SG & 3.62$\times10^{-3}$ \\
M92 & Bright HB    &  Faint HB & 6.79$\times10^{-4}$ \\
M92 & MR--SG    &  Bright HB & 1.33$\times10^{-4}$ \\
M92 & MR--SG    &  Faint HB & 0.980 \\
\enddata
\end{deluxetable}

\begin{figure*}
\epsscale{1.1}
\figurenum{11}
\plotone{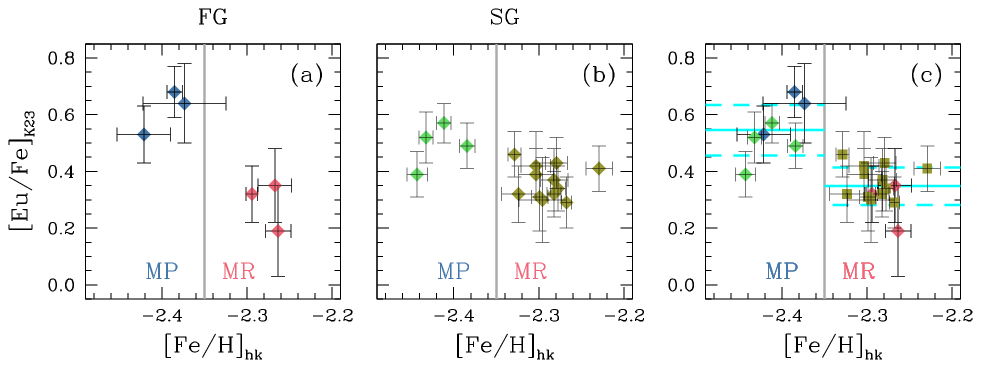}
\caption{
Plots of the \fehhk\ vs. [Eu/Fe] by \citet{kirby23} for the FG and SG RGB stars. The vertical gray lines denote the boundaries between the M92--MP and MR populations in our study \citep{lee23a}. (a) For the FG both from the M92--MP and MR populations. (b) For the SG RGB stars both from the MP and MR. Note that the MP--SG shown with the light green diamonds are classified as the AGB stars by \citet{kirby23}. (c) Combined plot using (a) and (b). The horizontal cyan solid and dashed lines denote the mean [Eu/Fe] and the standard deviations.}\label{fig:r}
\end{figure*}

\section{The [E\lowercase{u}/F\lowercase{e}] abundances in M92 as an evidence of a merger remnant}
As we mentioned above, \citet{kirby23} discovered the variations in the $r$-process neutron capture elemental abundances only in the M92 FG, while no such variation in the SG, which led them to suggest a minimum temporal separation of 0.8 Myr ($\sim$ several crossing times of M92) between the FG and SG. Here, we reinterpret their results in the context of our framework of M92 as a merger remnant of the two GCs.

In Figure~\ref{fig:r}, we show plots of our \fehhk\ versus [Eu/Fe] by \citet{kirby23} for the M92 RGB stars, where we classify individual RGB stars by their \fehhk\ (for the MP and MR), \cfech\ and \nfenh\ (for the FG and SG). We note that our FG and SG populations are about equivalent to those based on the [Na/Fe] and [Mg/Fe] abundances by \citet{kirby23}. We also note that all four MP--SG stars shown with light green diamonds in the panel (b) were classified as AGB stars by \citet{kirby23}. However, our multi--color photometry suggests that they are indeed MP--SG RGB stars. Our populational number ratio for the M92--MP population, $n$(FG):$n$(IG+SG) $\sim$ 49:41($\pm$4), is in agreement with the number ratio of the spectroscopic targets, $n$(MP--FG):$n$(MP--SG) = 3:4. Therefore, it should be reasonable to consider these four stars as MP--SG RGB stars.

Figure~\ref{fig:r}(a) shows a large hiatus between the [Eu/F] of the MP--FG and MR--FG, which led \citet{kirby23} concluded that the M92 FG stars show variation in the $r$-process neutron capture elemental abundances. In Table~\ref{tab:eu}, we show the mean [Eu/Fe] values and dispersions for individual populations. The [Eu/Fe] dispersion of the FG is about three times larger than that of the SG. However, we stress again that the MP and MR are independent populations with different abundances and structural parameters.

When we rearrange individual stars with their metallicity groups, the scatters in [Eu/Fe] become smaller as shown in Figure~\ref{fig:r}(c). The mean [Eu/Fe] values do not agree each other, with $\Delta$[Eu/Fe] = 0.198$\pm$0.038. Note that the difference is more than 5$\sigma$ levels. The figure also suggests that there is no [Eu/Fe] evolution within the MP and MR populations.

We believe that the different [Eu/Fe] abundances between the MP and MR populations provides a strong observational constraint on the true nature of M92 as the merger remnant. Our current result may suggest that they were independently formed out of chemically well mixed interstellar media.

\begin{deluxetable}{lcr}
\tablenum{5}
\tablecaption{Mean [Eu/Fe] values} \label{tab:eu}
\tablewidth{0pc}
\tablehead{\colhead{} & \colhead{$\langle$[Eu/Fe]$\rangle$} & \colhead{n}}
\startdata
FG & 0.452$\pm$0.073($\pm$0.178) &  6 \\
SG & 0.363$\pm$0.016($\pm$0.055) & 12 \\
\hline
MP & 0.546$\pm$0.034($\pm$0.089) &  7 \\
MR & 0.348$\pm$0.017($\pm$0.066) & 15 \\
\enddata
\end{deluxetable}

\section{SUMMARY}
We performed a comparative study of M30 and M92 using our own photometric system.
For M30, we obtained the mean photometric metallicity of \fehhk\ = $-$2.194$\pm$0.002 dex ($\sigma$ = 0.041), which is about 0.1 dex more metal--rich than M92 is. Our photometric metallicities are in good agreement with those in literatures \citep[e.g., 2010 version of][]{harris96}

We derived the photometric carbon and nitrogen abundances for both clusters. The mean primordial \cfech\ abundances are in good agreement in both clusters, although the \cfech\ dispersion of M92 slightly larger than that of M30, consistent with the fact that M92 has a more extended C-N anticorrelation with a more helium enhanced population. On the other hand, M30 appears to have a more enhanced primordial nitrogen abundance by \dnfe\ = 0.2 dex than M92 does. It is believed that M30 and M92 formed out of interstellar media experienced different chemical enrichment histories.

We investigated the metallicity and helium abundance differences between individual populations using our RGBB $V$ magnitudes, which sensitively depend on elemental abundances. We obtained the helium abundance difference of \dy\ = 0.025$\pm$0.016 between M30 FG and SG, which is in excellent agreement with that of \citet{milone18}, \dy$_{\rm max}$ = 0.022$\pm$0.010.

By assuming the same helium abundance, we obtained the metallicity difference of \dfeh\ = 0.200$\pm$0.051 dex between M92--MP and MR from the RGBB $V$ magnitudes, which is marginally in agreement with that of photometric metallicities, $\langle$\dfeh$\rangle$ = 0.119$\pm$0.004($\pm$0.061).

The most intriguing result is a large difference in the helium abundance between the M92 MR--FG and SG, \dy\ = 0.052$\pm$0.016. The lack of bright RGB stars in M92 MR--SG population also support the idea that it is indeed significantly enhanced in helium. It is believed that our result is  marginally in agreement with those by others \citep[e.g.,][]{milone18, vandenberg22, ziliotto23}.

We constructed evolutionary HB population synthesis models, finding that our metallicities and relative helium abundances satisfactorily reproduce the observed HB distributions for both clusters. A mild helium enhancement of \dy\ = 0.020--0.030 in the M92 MR--SG population failed to reproduce the extended BHB population with $V$ $\geq$ 16.1 mag in M92.

The CRDs may also suggested that the M92--MR--SG is significantly more helium enhanced than other populations and it is the progenitor of the faint HB.
The most spatially extended CRD of the M92--MR--SG is  due to their small stellar masses resulted from their fast stellar evolution with an significantly enhanced helium abundance by \dy\ $\sim$ 0.05. The CRDs of the M92--MR--SG and the faint HB are very similar with a $p$ value of 0.980 from a K--S test.
The fractions of the M92--MR--SG and the faint HB with respect to the total RGB and HB populations are also very similar, 0.18($\pm$0.02) and 0.20($\pm$0.03), respectively, strongly indicating that the faint HB stars with $V$ $\geq$ 16.1 mag are progeny of the M92--MR--SG RGB stars.

In our previous study \citep{lee23a}, we argued that M92 is a metal--complex GC, with a bimodal metallicity distribution and an atypical FG population, requiring significant magnesium and oxygen depletion. Here, we discussed a strong observational line of evidence of a large helium enhancement in MR--SG population. We also showed that M92 MP and MR populations have significantly different populational number ratios and spatial distributions. Furthermore, our reanalysis of [Eu/Fe] measurements by \citet{kirby23} suggested that the MP and MR populations are chemically separate populations, suggesting that both the MP and MR populations in M92 were most likely formed out of chemically well mixed interstellar media. Our results may suggest that M92 is not a typical globular cluster but a more complex system, perhaps a merger remnant of two GCs or a remnant nucleus of the metal--poor progenitor dwarf galaxy \citep[e.g., see][and references therein]{thomas20, lee23a}.

\begin{acknowledgements}
J.-W.\ Lee thanks Christopher Sneden for his comments on the early version of the manuscript and the anonymous referee for the useful suggestions. We acknowledge financial support from the Basic Science Research Program (grant No.\ 2019R1A2C2086290) through the National Research Foundation of Korea (NRF) and from the faculty research fund of Sejong University in 2022.
\end{acknowledgements}

\facilities{CTIO: 1.0 m (STA), WIYN: 0.9 m (HDI, S2KB), Gaia}
\software{\atlas\ \citep{kurucz11}, \moogscat\ \citep{moogscat}, \linemake\ \citep{linemake}}

\begin{figure*}
\epsscale{1.}
\figurenum{12}
\plotone{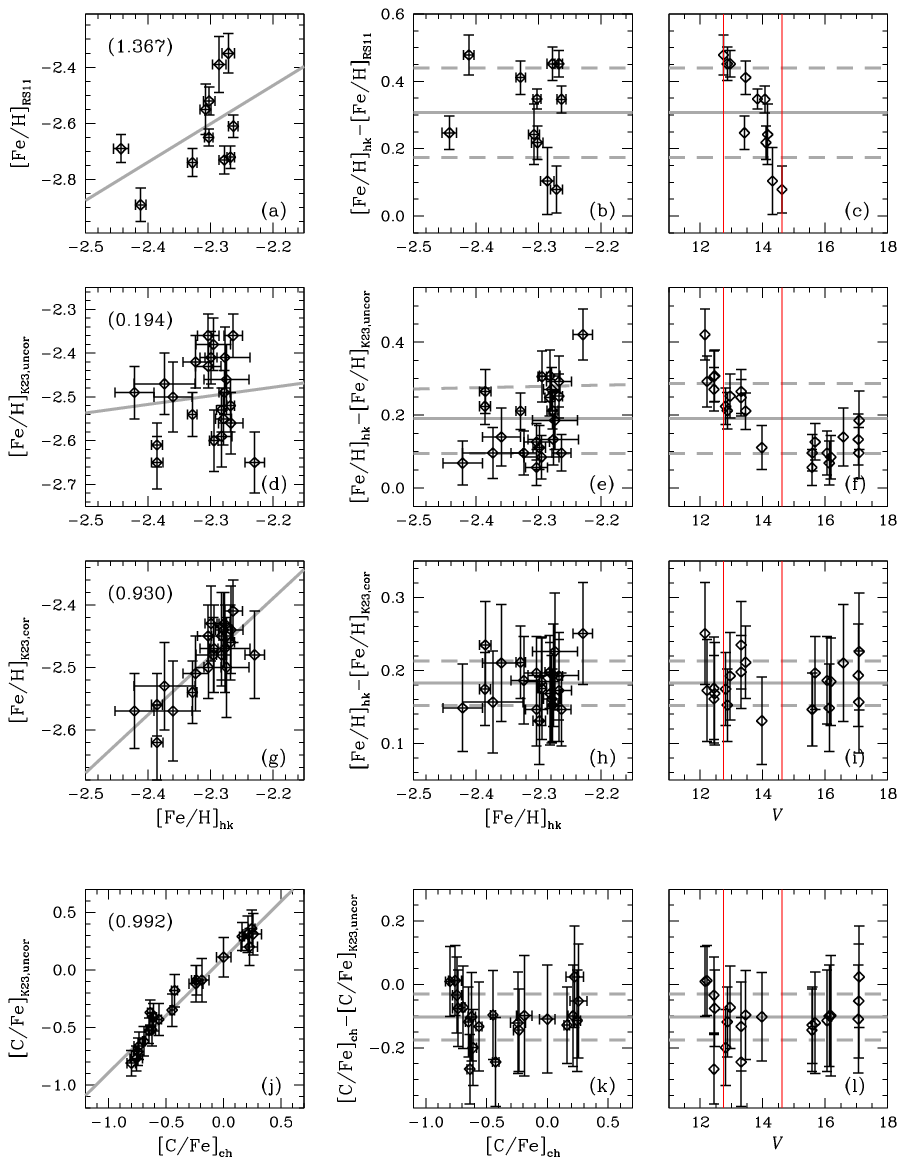}
\caption{
(a--c) Comparisons of our \fehhk\ versus \feh\ by \citet{roederer11}. In (a) the thick grey line denote the linear fit and the number in the parenthesis is the slope of the relation. In (b--c), the thick solid lines are the mean and the thick dashed grey lines denote the standard deviations ($\pm\sigma$). In (c), the red vertical lines denote the $V$ magnitude span of the RGB stars common in our study and that of \citet{roederer11}.
(d--f) Comparisons of our \fehhk\ versus the luminosity effect uncorrected \feh\ by \citet{kirby23} for RGB stars of our interest.
(g--i) Same as (d--f) but for the the luminosity effect uncorrected \feh.
(j--l) Comparisons of our \cfech\ versus the luminosity effect uncorrected \cfe\ by \citet{kirby23}.
}\label{fig:kirby}
\end{figure*}

\begin{figure*}
\epsscale{1.15}
\figurenum{13}
\plotone{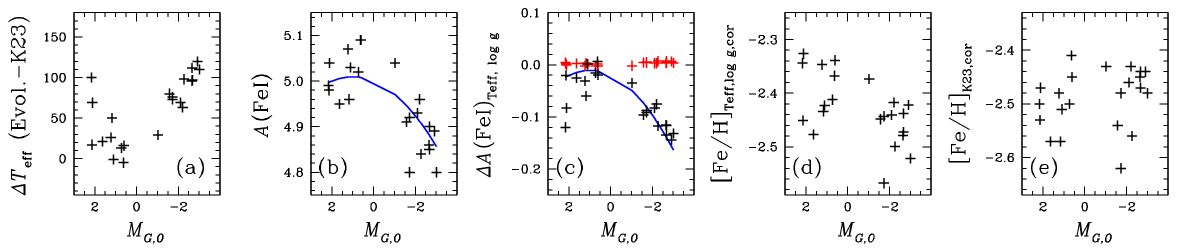}
\caption{
(a) Differences in effective temperatures between our evolutionary population synthesis models and those of \citet{kirby23}.
(b) Run of the uncorrected $A$(FeI) against $M_{G,0}$ of \citet{kirby23}. The blue solid line denotes the polynomial fit given Figure~4 of \citet{kirby23}.
(c) The black crosses denote the abundance corrections due to the difference in the effective temperature in (a), while the red crosses those due to difference in the surface gravity. The blue solid line is for the polynomial fit in (b) minus 5.02.
(d) The differential temperature corrected [Fe/H].
(e) The luminosity-based gradient corrected [Fe/H] by \citet{kirby23}.
}\label{fig:gradient}
\end{figure*}

\begin{figure*}
\epsscale{0.93}
\figurenum{14}
\plotone{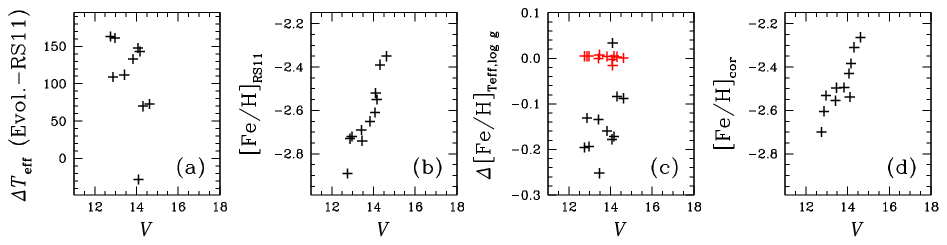}
\caption{
(a) Differences in effective temperatures between our evolutionary population synthesis models and those of \citet{roederer11}.
(b) Run of the uncorrected \feh\ against $V$ magnitude.
(c) The abundance corrections based on the panel (a). The red crosses denote the corrections due to different adopted surface gravity, which does not greatly affect the results.
(d) The differential temperature corrected [Fe/H].
}\label{fig:rs11}
\end{figure*}

\appendix
\section{Comparisons of our [F\lowercase{e}/H]$_{\lowercase{hk}}$ and [C/F\lowercase{e}]$_{\lowercase{ch}}$ of M92 RGB stars to the [F\lowercase{e}/H] and [C/F\lowercase{e}] by Kirby, Ji, \& Kovalev}\label{ap:m92}
In our previous study for M92 \citep{lee23a}, we showed comparisons of our \fehhk\ measurements of the M92 RGB stars to those of the differential spectroscopic analysis by \citet{roederer11}, finding a large mean metallicity offset, $\langle$\fehhk\ $-$ \feh$_{\rm RS11}\rangle$ = 0.314$\pm$0.039 dex ($\sigma$ = 0.128 dex), and a strong luminosity or temperature gradient.\footnote{In Figure~3 of \citet{lee23a}, \dfeh\ denotes the residuals around the fitted lines, not the differences between our \fehhk\ and \feh\ by \citet{roederer11}, \fehhk\ $-$ \feh$_{\rm RS11}$. We showed the residuals around the fitted line to demonstrate the luminosity effect in the \feh\ measurements by \citet{roederer11}.}

In Figure~\ref{fig:kirby}(a--c), we show comparisons of our \fehhk\ versus \feh\ by \citet{roederer11}. We note a strong gradient in the metallicity difference between our measurements and those of \citet{roederer11} against the $V$ magnitude in the panel (c), in the sense that the \feh\ values of \citet{roederer11} become significantly smaller with increasing luminosity. As shown in Figure~\ref{fig:abund}, our \fehhk\ does not show any gradient against the $V$ magnitude and, therefore, the gradient in (c) should be arisen from the \feh\ measurements by \citet{roederer11}.

Recently, \citet{kirby23} performed precision high resolution spectroscopy of M92 stars, spanning from the RGB-tip to the subgiant, with much better spectra qualities than those used by \citet{roederer11}. In Figure~\ref{fig:kirby}(d--f), we show comparison of our \fehhk\ versus uncorrected \feh\ measurements by \citet[][see their Table~5]{kirby23}. The correlation between the two datasets does not look good with the Pearson correlation coefficient of 0.106. The strong abundance gradient against luminosity persists in the panel (f), although the mean difference is somewhat smaller than that from \citet{roederer11},  $\langle$\fehhk\ $-$ \feh$_{\rm K23, uncor}\rangle$ = 0.191$\pm$0.020 dex ($\sigma$ = 0.095 dex).

\citet{kirby23} noticed considerable elemental abundance trends with the stellar luminosity even after applying non-LTE corrections for some elements. In order to remove the unexplained trend, they interpreted these trends as artifacts of their measurements. They derived polynomial fits of elemental abundances against the $M_{G,0}$ magnitude and they showed their results in their Figure~4. In Figure~\ref{fig:kirby}(g--i),  we show comparison of our \fehhk\ versus the luminosity-based gradient corrected \feh\ measurements by \citet{kirby23}. As shown in the panel (g), the correlation between the two datasets is excellent with a correlation coefficient of 0.824. The panel (i) clearly showed that the elemental abundance gradient against luminosity disappears. We find that $\langle$\fehhk\ $-$ \feh$_{\rm K23, cor}\rangle$ = 0.183$\pm$0.007 dex ($\sigma$ = 0.031 dex). Note that the dispersion is also significantly improved, which is what we demonstrated in our previous study in comparison with the results by \citet{roederer11}. Below, we will argue that the inappropriate temperature scales are responsible for this discrepancy, at least a part.

\citet{kirby23} also measured carbon abundances by employing spectral synthesis of the CH G-band features. As kindly noted by Dr.\ Kirby that the [C/Fe] values in their Table~6 are based on the uncorrected [C/H] and the luminosity-based gradient corrected [FeI/H]. When we compare our \cfech\ with the [C/Fe] values in their Table~6, a strong gradient in the difference in carbon abundances (\cfech\ $-$ \cfe) against the luminosity, similar to our Figure~\ref{fig:kirby}(f), can be seen.
As recommended by Dr.\ Kirby, we derived [C/Fe] based on their uncorrected [C/H] and [Fe/H] measurements, [C/Fe]$_{\rm K23, uncor}$, and we show our comparisons in Figure~\ref{fig:kirby}(j--l). Our \cfech\ is nicely correlated with the spectroscopic measurements by \citet{kirby23} without any perceptible trend against luminosity, with the Pearson correlation coefficient of 0.981.
The mean difference between the two measurements is $\langle$\cfech\ $-$ \cfe$_{\rm K23, uncor}\rangle$ = $-0.103\pm0.016$ dex ($\sigma$ = 0.073 dex).

As we will discuss below, the absence of any gradient against luminosity in \cfech\ $-$ \cfe\ shown in Figure~\ref{fig:kirby}(l) can be explained with variations in effective temperature of the input atmosphere models. Qualitatively, for example, in metal--poor stars, as effective temperature increases the number densities of CH and Fe I both decrease while at the same time the H$^{-}$ continuum opacity increases due to increasing electron density from ionizing metals. As a consequence, the correlation between uncorrected [C/H] and [Fe/H] with the effective temperature should exist. Therefore any artificial trends from inappropriate input effective temperature should be canceled out in the uncorrected [C/Fe]. This is the reason why we do not see any gradient in our Figure~\ref{fig:kirby}(l).
We simulate this with \moogscat\ for the star V--45, which appears to have the best spectrum quality among other stars. Using the equivalent width measurements by \citet{kirby23}, we obtained $\delta$[Fe/H]/$\delta$\teff(+100K) = 0.12 dex K$^{-1}$. Since \citet{kirby23} relied on the synthetic spectra for their carbon abundance measurements, we generated synthetic spectra around the CH G band region, and we obtained $\delta$[C/H]/$\delta$\teff(+100K) = 0.15 dex K$^{-1}$, confirming our qualitative reasoning that the temperature dependency of the CH G band absorption strength is about the similar to that of the neutral iron atoms.

We conclude that our \cfech\ measurements are in excellent agreement with those from the high resolution spectral syntheses of the CH G band. It should not be surprising that both results are based on the same spectral features and analyzing tools, but different approaches; photometric versus spectroscopic.

\section{Effective Temperatures: Origin of the elemental gradient against luminosity in high resolution spectroscopic study}\label{ap:teff}
As mentioned above, \citet{kirby23} noticed considerable elemental abundance gradients against the stellar luminosity. They applied non-LTE corrections for some elements, but these gradients persist. This is at odd. We used the similar LTE approach with the same tools as those adopted by \citet{kirby23}, but our \fehhk\ does not show any perceptible gradient with the stellar luminosity, while the results of \citet{kirby23} show considerable quadratic trends in elemental abundances against luminosity. The major difference between the two studies is the adopted input stellar parameters, in particular the effective temperature. We relied on the evolutionary effective temperatures, while \citet{kirby23} used those estimated from the color--temperature relations by \citet{mucciarelli21}. We argue that their additional luminosity-based gradient is arisen from their inappropriate temperature scales.

In Figure~\ref{fig:gradient}(a), we compare our evolutionary \teff\ and photometric \teff\ adopted by \citet{kirby23}. As we discussed above, our evolutionary population synthesis models are based on the latest BaSTI \citep{basti21}. We emphasize again that our \fehhk\ measurements do not show any gradient against the $V$ magnitude as in Figure~\ref{fig:abund}, strongly suggesting that our input parameters, such as the effective temperature and surface gravity, are self-consistent. Below and above $M_{\rm G,o}$ $\sim$ 0.0 mag, the effective temperatures adopted by \citet{kirby23} are significantly smaller than those from our evolutionary models. At given observed equivalent widths of neutral iron atoms, lowering input effective temperatures in the LTE analysis will result in lowering output [Fe/H] values in RGB stars.

In Figure~\ref{fig:gradient}(b), we show the run of uncorrected $A$(FeI) by \citet{kirby23} against the $M_{G,0}$ magnitude, showing a considerable curvature. Also shown is their polynomial fit to the iron abundances to remove the trend.

In Figure~\ref{fig:gradient}(c), we show the amount of the iron abundances that were under- or over-estimated due to the temperature differences in the panel (a). To calculate the correction value, we adopted $\delta$[Fe/H]/$\delta$\teff(+100K) = 0.12 dex K$^{-1}$ from our result above. We also calculated those from the different surface gravity, but their contributions are insignificant as shown in the panel (c). In the figure, we show the modified polynomial fit, which is the polynomial fit by \citet{kirby23} minus 5.02. In the panel (c), the modified polynomial fit nicely traces the [Fe/H] correction values due to temperature differences of individual stars. We applied these correction values to individual stars as shown in the panel (d), and we obtained the mean temperature and surface gravity difference corrected \feh\ value of $\langle$\feh$_{\rm Teff, log g, cor}\rangle$ = $-2.429\pm0.013$ dex ($\sigma$ = 0.061 dex). Note that our mean \fehhk\ value from these common stars is $\langle$\fehhk$\rangle$ = $-2.309\pm0.010$ dex ($\sigma$ = 0.047 dex) and the discrepancy in the metallicity difference between the two studies is slightly mitigated. Our new [Fe/H] value is about 0.06 dex larger than that from the luminosity-based polynomial fit by \citet{kirby23}, $\langle$\feh$_{\rm Teff,cor}\rangle$ = $-2.491\pm0.011$ dex ($\sigma$ = 0.054 dex), as shown in (e).

We conclude that the luminosity-based elemental abundance trends seen by \citet{kirby23} are due to the inappropriate temperature scale adopted by them, which strongly indicating that the color--temperature relations by \citet{mucciarelli21} fail to work, at least for M92. We note that the distance modulus and interstellar reddening value are well known for M92 and the quadratic trend around $M_{G,0}$ cannot be explained by them.

We applied the same procedures for the \feh\ measurements by \citet{roederer11}. Again, the photometric effective temperatures adopted by \citet{roederer11} are cooler than our evolutionary effective temperatures, $\langle$\teff$\rangle$ = $118\pm18$ K ($\sigma$ = 60 K, 11 stars). The temperature correction will increase their mean metallicity by $\langle$\feh$_{\rm Teff,cor} -$\feh$_{\rm uncor}\rangle$ = $0.141\pm0.022$ dex ($\sigma$ = 0.072 dex) but it does not completely remove the trend against luminosity as shown in Figure~\ref{fig:rs11}(d), which may require additional explanation.\footnote{The referee also confirmed a strong trend of metallicity from spectroscopy by \citet{roederer11} as a function of magnitude and temperature along the RGB in M92, with warmer and fainter stars showing higher metallicities. The referee pointed out that this trend reflects a similar trend as a function of S/N and may then be driven by (or, at least, due in part to) the quality of spectra. This could be maybe the additional explanation invoked by us, beside the lowering of [Fe/H] values due to the lowering input effective temperatures.}

\begin{figure*}
\epsscale{.9}
\figurenum{15}
\plotone{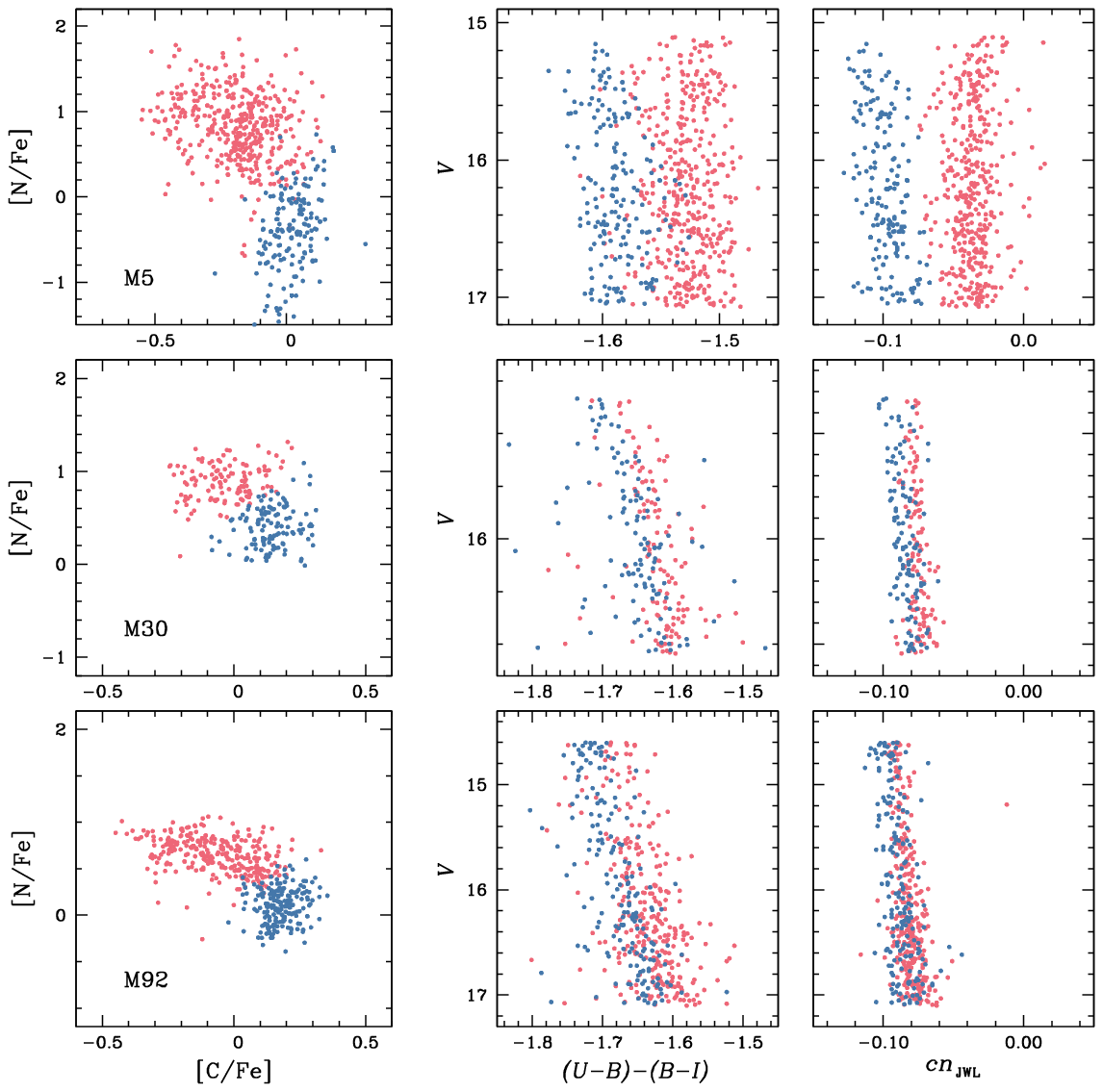}
\caption{
Plots of [C/Fe] vs.\ [N/Fe], \cubi(= $(U-B)-(B-I)$) vs.\ $V$, and \cnjwl\ vs.\ $V$ of RGB stars fainter than their RGBBs in M5, M30, and M92. We use the $UBI$ photometry by \citet{pbs19}. The blue and red dots are for the FG and SG populations classified based on the photometric elemental abundances in our previous study \citep[][for M5]{lee21b} and our current study (for M30 and M92). Our results show that the \cubi\ is not capable of distinguishing MPs in metal--poor GCs.}\label{fig:cubi}
\end{figure*}

\begin{figure*}
\epsscale{.9}
\figurenum{16}
\plotone{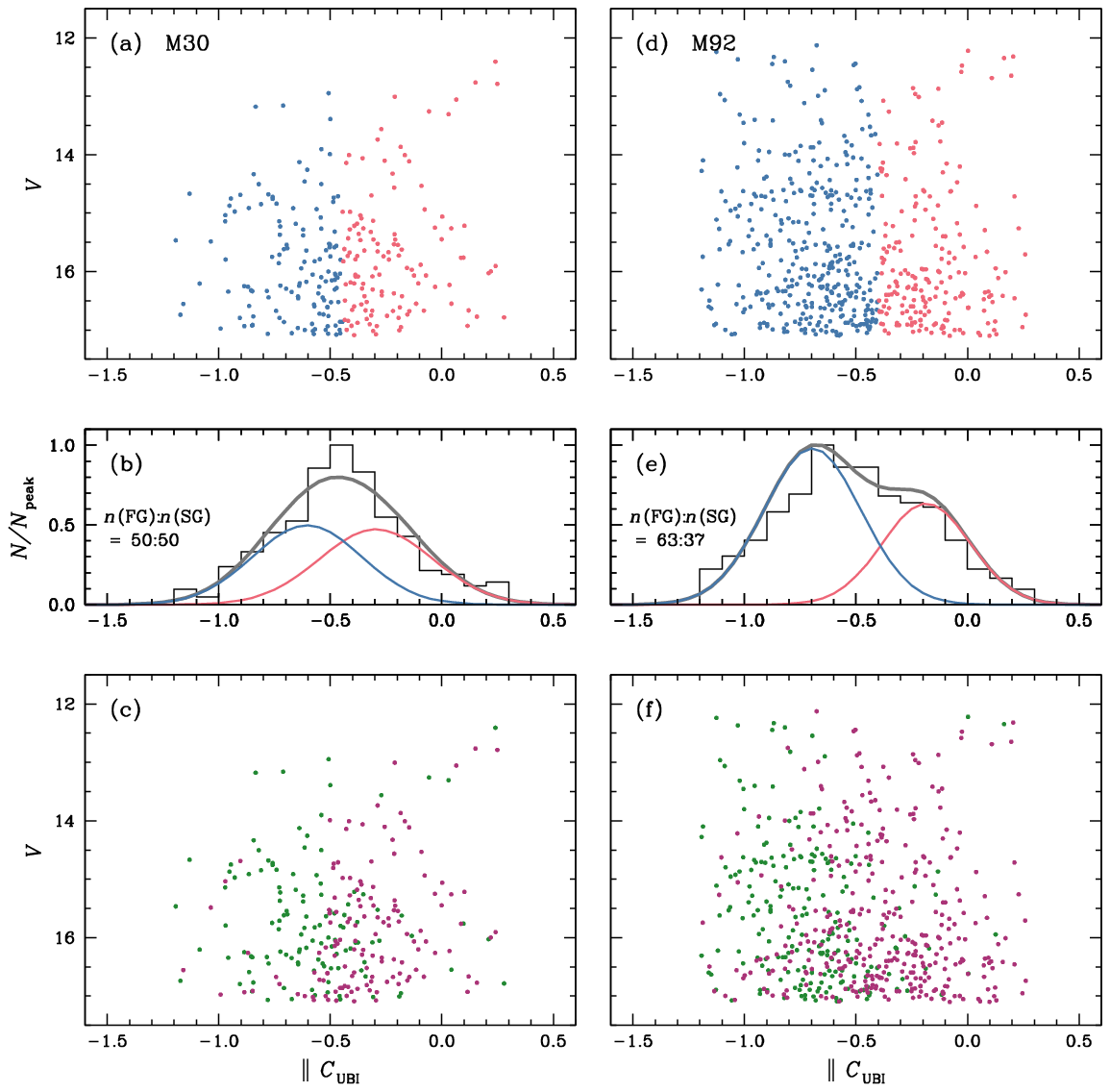}
\caption{
(a) A plot \pcubi\ vs.\ $V$ of RGB stars in M30. The blue dots are for the FG and the red dots are for the SG based on the \pcubi\ distribution.
(b) A histogram for the \pcubi\ distribution. The blue and red solid lines denote the FG and SG returned from our EM analysis of the \pcubi\ distribution, while the gray solid line is for the total RGB population.
(c)  A plot \pcubi\ vs.\ $V$ of the FG (green) and SG (purple) RGB stars in M30 from our current study, showing a gradual transition from one population to the other.
(d)--(f) Same as (a)--(c) but for M92
}\label{fig:emcubi}
\end{figure*}

\section{On the use of the $C_{\mathrm UBI}$ index in the metal-poor regimes}\label{ap:cubi}
It seems very attractive to make use of the broadband $UBI$ photometry and the relevant pseudo color index, \cubi\ = $(U-B)-(B-I)$, in the GC MSP study, since a huge amount of $UBI$ photometric data has been already accumulated and publicly available over the years, such as nicely presented by \citet{pbs19}. However, it is a great tragedy that many astronomers claimed their own results for the GC MSP characteristics, such as populational number ratios, CRDs, and etc., by simply employing the \cubi\ without any careful validation.

In our previous study \citep{lee19a}, we demonstrated that the \cubi\ index in the GC MSP study is not reliable due to the inherited trait of the broadband photometry. Using three intermediate metallicity GCs, M5, M3, and NGC\,6752, we compared performances of populational taggings from our \cnjwl\ and the \cubi\ indices, finding that the results from the \cubi\ are easily in error. We proposed further that the \cubi\ should not be used in the metal-poor domain \citep[e.g., see][]{lee23a}. It is because the \cubi\ makes great use of the CN absorption band features as shown in Figure~1 of \citet{lee19a}. However, due to a double--metal molecule nature of CN, its absorption strength rapidly decline with decreasing metallicity and CN is almost absent in the metal-poor RGB stars, such as in M30 and M92, especially the CN red system resided in the $I$ passband.

In Figure~\ref{fig:cubi}, we show plots of [C/Fe] vs. [N/Fe], $V$ vs. \cubi, and $V$ vs. \cnjwl\ of RGB stars fainter than their RGBBs in M5 \citep{lee21b}, M30, and M92 \citep{lee17, lee19a, lee23a}. In the figure, we classified individual populations following our tagging scheme described in \S\ref{sec:tagging}. For M5 and M92, the SG includes both the IG and SG for M92 and the IG and the Extreme population for M5 \citep{lee21b}. For the \cubi\ index, we used the $UBI$ photometry by \citet{pbs19}. In the figure, we distinguish the FG and SG only for clarity. In the intermediate metallicity GC M5, our \cnjwl\ index provides two discrete populations. On the other hand, the boundaries of the two populations are overlapped in the \cubi\ index and caution should be used to separate the correct population in the mixed zone.

For metal--poor GCs M30 and M92, both the \cnjwl\ and \cubi\ are not capable of distinguishing individual populations. This is especially true for our \cnjwl, since it mainly relies on the CN band absorption feature at \cnwave\AA. Variations in the nitrogen and carbon abundances among MSPs can produce some gradual change of \cubi, but the discrete populational separation cannot be seen. Figure~\ref{fig:emcubi} shows the situation more clearly. We normalized the \cubi\ indices for M30 and M92, \pcubi, and we performed an EM analysis on their \pcubi\ distributions, finding the populational number ratio of \nrgb\ = 50:50($\pm$5) for M30 and \nrgb\ = 63:37($\pm$4) for M92. This populational number ratio of M30 returned from the \pcubi\ is apparently exactly the same value as that from the (\dcfenfe) distribution, 50:50($\pm$5), of our current study. However, the populational tagging of the individual stars tells that it is a coincidence. As Figure~\ref{fig:emcubi}(c) and (f) show, the FG stars classified from our photometric abundance overlap with the SG stars, i.e., a gradual transition from one population to the other in metal-poor GCs. Therefore, the populational taggings cannot be done correctly based on the \cubi\ alone. For this reason, we performed populational tagging based on our \pcfe\ and \pnfe\ domain, which seems to work excellently.

\begin{figure*}
\epsscale{1.1}
\figurenum{17}
\plotone{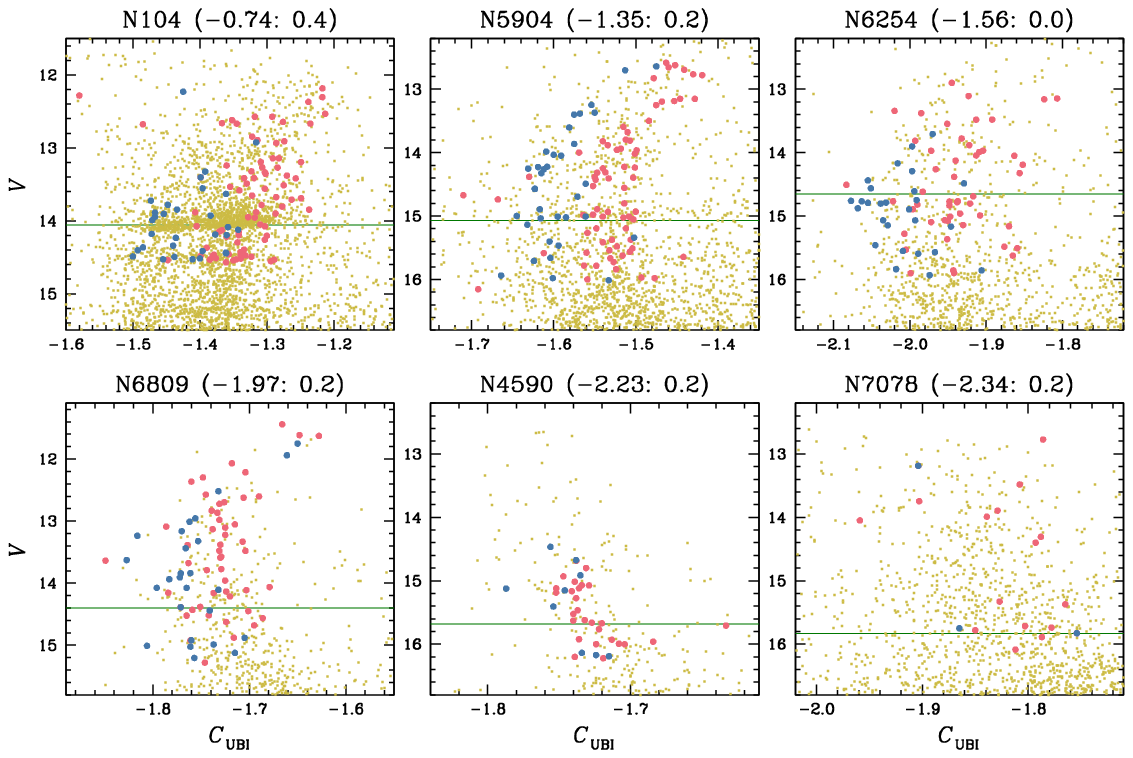}
\caption{Plots \cubi(= $(U-B)-(B-I)$) vs.\ $V$ of RGB stars in six GCs. The blue dots are for the FG and the red dots are for the SG based on [Na/Fe] of \citet{carretta09}. Yellow dots are for all stars by \citet{pbs19}. The horizontal dark green lines denote the mean $V$ magnitude of the HB stars in each GC. On the top of each box, we show the name, [Fe/H], and [Na/Fe] for the FS and SG separation.}\label{fig:gcs}
\end{figure*}

Finally, we show plots of $V$ versus \cubi\ of six representative GCs in Figure~\ref{fig:gcs}. We show the FG and SG RGB stars classified with the [Na/Fe] measurements by \citet{carretta09}; [Na/Fe] $\leq$ 0.2 dex for the FG and [Na/Fe] > 0.2 dex for the SG. For NGC\,104 and NGC\,6254, we adopted slightly different [Na/Fe] values to secure proper numbers in the FG. We note that our adopted [Na/Fe] does not affect our main issue addressing here. For the metal--rich GC NGC\,104 and the intermediate metallicity GC NGC\,5904, the \cubi\ index may be a barometer of the populational tagging of the RGB stars. However, when it comes to metal--poor GCs, the \cubi\ becomes a hopelessly terrible index, due to the reasons that we mentioned above. Figure~\ref{fig:gcs} clearly shows severe confusions in populational taggings with the \cubi\ emerge in the whole RGB luminosity level in metal--poor regimes, a strong evidence of an intrinsic drawback of the \cubi. We stress again that the populational taggings cannot be done with the \cubi\ alone in the metal--poor domain, and more dimensions are definitely needed to do the job correctly.

\end{document}